\documentclass[prd,twocolumn,superscriptaddress,floatfix,amsmath,amssymb]{revtex4}

\usepackage{amsmath,amssymb,amsthm,amsfonts}
\usepackage{verbatim}

\usepackage{listings}
\usepackage{graphicx}
\usepackage{verbatim}
\usepackage{color}
\usepackage{hyperref}

\newcommand{\be}{\begin{equation}}
\newcommand{\ee}{\end{equation}}
\newcommand{\ba}{\begin{eqnarray}}
\newcommand{\ea}{\end{eqnarray}}
\usepackage{multirow}

\newcommand{\beq}{\begin{equation}}
\newcommand{\eeq}{\end{equation}}
\newcommand{\beqa}{\begin{eqnarray}}
\newcommand{\eeqa}{\end{eqnarray}}

\definecolor{forestgreen}{RGB}{34,139,34}
\usepackage{MnSymbol}
\lstdefinelanguage{Maxima}{
  keywords={addrow,addcol,zeromatrix,ident,augcoefmatrix,ratsubst,diff,ev,tex,%
    with_stdout,nouns,express,depends,load,submatrix,div,grad,curl,%
    rootscontract,solve,part,assume,sqrt,integrate,abs,inf,exp,reveal,read,if,%
    then,expand,print,array,listarray,subst,sequal,string,length,hipow,for,%
    thru,do,gamma,else},
  sensitive=true,
  morestring=[b]",
  stringstyle=\color{forestgreen},
  comment=[n][\itshape]{/*}{*/}
}

\begin{document}

\title{Mass and Thermodynamic Volume in Lifshitz Spacetimes}

\author{W. G. Brenna}
\email{wbrenna@uwaterloo.ca}
\affiliation{Department of Physics and Astronomy, University of Waterloo, Waterloo, Ontario N2L 3G1, Canada}
\author{Robert B. Mann}
\email{rbmann@uwaterloo.ca}
\affiliation{Department of Physics and Astronomy, University of Waterloo, Waterloo, Ontario N2L 3G1, Canada}
\author{Miok Park}
\email{miokpark@kias.re.kr}
\affiliation{School of Physics, Korea Institute for Advanced Study, Seoul 130-722, Korea}

\begin{abstract}
We examine the concept of black hole thermodynamic volume and its consistency with
 thermodynamic mass  in spacetimes that are not asymptotically flat but instead have anisotropic Lifshitz scaling symmetry. We find that the generalized
 Smarr relation in anti de Sitter space -- extended to include a pressure-volume term -- holds here as well,
 and that  there exists a
definition of thermodynamic mass and thermodynamic volume that satisfy both this relation and the
$1^{st}$ law of thermodynamics.
We compare the thermodynamic mass with other known
quantities such as Arnowitt-Deser-Misner, Brown-York and Hollands-Ishibashi-Marolf masses.
We also conjecture methods for obtaining a thermodynamic mass where there is ambiguity
due to the cosmological constant lengthscale depending on the horizon radius lengthscale.
\end{abstract}

\maketitle

\section{Background}

Gauge-gravity duality remains a subject of considerable interest, in large part because of the insights
it yields into quantum gravity.  Asymptotically anti de Sitter (AdS) spacetime
admits a strongly coupled gauge theory
description at its boundary via a holographic dictionary.
It is straightforward to define thermodynamic equilibrium in this case, in turn giving rise to thermal radiation/large AdS black hole phase transitions \cite{Hawking1983}.

An  interesting  development in this subject that has been the subject of much current interest is the proposal that the  mass of an AdS black hole can be understood as the enthalpy of  spacetime \cite{Kastor2009}.  This notion
emerges from regarding  the cosmological constant $\Lambda$  as a thermodynamic variable
\cite{Creighton1995} analogous to pressure in the first law \cite{Caldarelli2000, Kastor2009, Dolan2010, Dolan2011a, Dolan2011, Dolan2012, Cvetic2010, Larranaga2011, Larranaga2012,
Gibbons2012, Kubiznak2012, Gunasekaran2012, Belhaj2012,  Lu2012, Smailagic2012, Hendi2012}, along with a notion
of conjugate volume \cite{Kastor2009,Gibbons2012}.
 A complete analogy between 4-dimensional
Reissner-Nordstr\"om AdS black holes and the Van der Waals liquid--gas system can be shown to hold  \cite{Kubiznak2012}.  The  critical exponents are the same as those in the Van der Waals system, modifying previous considerations that emerged from earlier studies \cite{Chamblin1999a,Chamblin1999, Cvetic1999, Cvetic1999a} of the duality description. Intensive investigation in a broad variety of contexts \cite{Dolan2013, Dolan2013a,
Zou2013,Zou2014,Ma2013, Ma2014, Wei2014,Mo2014, Mo2014a, Mo2014b,Zhang2014, Kubiznak2014,Liu2014, Liu2014a, Johnson2014a} has led to the discovery of  a variety of new thermodynamic phenomena for both AdS and de Sitter \cite{Dolan2013b} black holes, including
the existence of reentrant phase transitions in  Born-Infeld \cite{Gunasekaran2012} and rotating \cite{Altamirano2013} black holes, the existence of a tricritical point in rotating black holes analogous to the triple point in water \cite{Altamirano2013},  a new type of thermodynamic criticality in
the higher-curvature case \cite{Frassino2014}, and the notion of a holographic heat engine
  \cite{Johnson2014a,MacDonald2014}.  Indeed, the thermal radiation/large AdS black hole phase transition \cite{Hawking1983} can be understood as a solid/liquid phase transition from this perspective \cite{Kubiznak2014}.

Here we begin the first study of extended thermodynamics in the context of Lifshitz duality.
Motivated by the hope of obtaining a duality between condensed matter physics with quantum criticality, the  anisotropic  scaling properties of these systems imply from gauge-gravity duality that the  bulk spacetime geometry likewise asymptotes to a spacetime with the same scaling properties  \cite{Kachru2008}.  Known as  Lifshitz spacetimes, they remain a subject of intensive study \cite{Alvarez2014,Ghanaatian2014,Liu2014b,Park2014}.

We seek to understand the thermodynamics of Lifshitz  black holes \cite{Danielsson2009,Mann2009,Bertoldi2009}  in the context of extended phase  space, particularly the notion of mass as enthalpy \cite{Kubiznak2012}.  Mass, a difficult concept to define in general relativity, is even more problematic when higher-curvature and/or differing asymptotics are incorporated. There are a number of competing  definitions that often agree in specific cases; for example, the Arnowitt-Deser-Misner (ADM) and Komar masses agree for stationary asymptotically flat spacetimes \cite{Ashtekar1979}.
Other definitions of mass have had
utility in various scenarios, including the
 Abbott-Deser-Tekin (ADT) mass, which applies to non-asymptotically-flat spacetimes \cite{Abbott1982};
the Wald formula, which yields a mass via the $1^{st}$-law \cite{Wald1993,Iyer1994};
counterterm methods \cite{Henningson1998,Balasubramanian1999,Mann1999,Skenderis2001};
the quasilocal Brown-York definition \cite{Brown1993}; and other masses \cite{Hollands2005,Wang2008a}, which
use charges appearing from various boundary stress-tensors.

Currently there exists
some disagreement over the correct mass to use for asymptotically Lifshitz  spacetimes. For example, \cite{Devecioglu2010}
proposed a mass that was later found not to satisfy the first law of thermodynamics \cite{Gim2014}.
Quasilocal formalisms \cite{Brown1993} have been successfully used in first-law driven approaches \cite{Gim2014}, but
debate exists over which quasilocal mass to use; for example, whether the
Brown-York mass is more or less appropriate than the Hollands-Ishibashi-Marolf mass \cite{Mann2011a}.
A final criticism of some quasilocal masses is that they are often not gauge invariant, and a technique for obtaining
a gauge invariant quasilocal mass has been put forward by Wang and Yau \cite{Wang2008a}.

We seek to understand if and how mass can be understood as enthalpy for
Lifshitz space times,  generalizing the AdS notions of pressure and conjugate thermodynamic volume to this setting.
Our task is, in part, to obtain a ``thermodynamic mass'' - a mass for Lifshitz black hole spacetimes that is consistent with both the first law of thermodynamics (i.e. consistent with
 standard definitions of temperature via Wick rotation and entropy \cite{Iyer1994}) as well as the more general Smarr relation
 (an integrated first law) that gives mass in terms of temperature, entropy \cite{Smarr1973},
and (more recently for AdS spacetimes), pressure and volume \cite{Kastor2009}.  Indeed, in asymptotically AdS space times, the requirement of a consistent Smarr formula necessarily entails
 inclusion of the  cosmological constant from both scaling  \cite{Caldarelli2000, Dolan2010} and geometric \cite{Kastor2009} considerations, leading to the pressure/volume interpretation noted above.

 One of the features of the Lifshitz class of black holes is that their
 asymptotic structure necessarily causes the (negative) cosmological constant to
 become dependent on the matter couplings in the theory, making the
 distinction of thermodynamic pressure less than clear. A number of
 different Smarr-like relations for Lifshitz black holes have appeared
 \cite{Bertoldi2009a,Dehghani2010,Liu2014b,Berglund2011,Dehghani2011c,Dehghani2013a, Way2012}, none making any reference to any pressure-volume terms that naturally appear in the AdS case \cite{Creighton1995,Caldarelli2000, Kastor2009, Dolan2010}.
For asymptotically Lifshitz spacetimes we expect the requirement of a consistent thermodynamic mass to yield a corresponding generalization of the Smarr relation, providing additional guidance in selecting an appropriate mass for these spacetimes.  We will find that such a generalized Smarr relation exists and is the same as the AdS case, with previous Smarr relations being recovered as special cases. In so doing we find that pressure retains the same interpretation it has in the AdS case, and the notion of thermodynamic volume is thereby extended to the Lifshitz setting.

The general asymptotically Lifshitz spherically symmetric metric ansatz is
\begin{equation}
ds^2 = -\left(\frac{r}{l}\right)^{2z}f(r) dt^2 + \frac{l^2 dr^2}{g(r)  r^2} + r^2 d\Omega_k^2
\label{eqn:metric}
\end{equation}
where $z \neq 1$ and $d\Omega^2_k$ is the metric of a hypersurface consisting of the  {$D-2$} remaining dimensions
(whose coordinates will be symbolically denoted by $x^i$). Note that the anisotropic scaling property
\begin{equation}
t \rightarrow \lambda^z t, \hspace{5mm} r \rightarrow \lambda^{-1} r, \hspace{5mm} x^i \rightarrow \lambda x^i
\end{equation}
holds for (\ref{eqn:metric}) provided  $f$ and $g$ both approach unity for large $r$; this scaling property is essential for generalized gauge-gravity duality.   When $z=1$  isotropic scaling is restored, leading to the AdS spacetime. If $f=g=1$
the spacetime is that of pure Lifshitz spacetime, and is generally regarded as playing a role in generalized
gauge-gravity duality similar to that of the AdS spacetime in the AdS/CFT correspondence \cite{Kachru2008}, though this interpretation has a number
of difficulties \cite{Copsey2010}.

We shall examine several  actions of the form
\begin{align}
	\mathcal{I} = \frac{1}{\kappa} \int & d^D x \sqrt{-g} \left( R - 2 \Lambda \right. \nonumber \\
	&+ \left. \mathcal{F}(R,R_{\mu \nu}, R_{\mu \nu \alpha \beta}) + \mathcal{G}(B_i,H_i) \right)
	\label{eqn:actiontemplate}
\end{align}
whose field equations yield solutions of the form (\ref{eqn:metric}). Here
 $\mathcal{F}$ is some polynomial function of higher curvature terms and
$\mathcal{G}(B_i,H_i)$ is some function of a set of vector fields $B_i$
and their respective field strengths $H_i$.

Solutions to the associated field equations have been relatively well studied in a number of contexts.
Typically the action is modified from general relativity by adding a Proca field in order to produce
the asymptotics necessary for the Lifshitz symmetry \cite{Copsey2010}; however, tuned higher-curvature terms can also
be used to this effect. Because of the unusual asymptotics (in the case of the Proca field, the field
potential is finite at infinity), much of the work on Lifshitz symmetric black holes is numeric;
nonetheless, some exact solutions have been found \cite{Mann2009,Bertoldi2009,Gim2014}.

We will first introduce the relevant equations for our method of obtaining a mass, followed by a
demonstration of the computation of mass for some  exact black hole solutions in various spacetimes.
In asymptotically flat spacetimes, when the Smarr relation  for spinless neutral black holes is written $(D-3) M = (D-2) TS$,
it is apparent that knowing $T$ and $S$ for the black hole will immediately yield knowledge
of the mass. However, when a pressure-volume term is added in the asymptotically AdS case,
this is less clear, since thermodynamic volume is not necessarily known \textit{a priori}.
One can turn to the $1^{st}$ law of thermodynamics to obtain another equation in an attempt
to find the mass, but it is not obvious whether a solution exists since
the problem is no longer linear,  unless the assumption is made that
the entropy and volume are independently related quantities.
We will show that without requiring independence of all thermodynamic
variables, the number of lengthscales for
the black hole system can sufficiently simplify the set of equations
to the point where the thermodynamic mass and volume can be solved for.
We will also see that without assuming independence of thermodynamic
variables on the lengthscales, 
this method is consistent with the technique of integrating $dM = TdS$
over the horizon radius $r_h$ to obtain a mass \cite{Cai1999,Cai2002},
in cases where the thermodynamic variables do turn out to be independent.

In summary, we shall see that constraining the mass
via the Smarr relation and $1^{st}$ law is a surprisingly strong
restriction; in many cases this mass is forced to agree with the ADM mass under these
and a few small assumptions.  We shall then employ this approach to  examine the masses of various black hole spacetimes 
for which no methods to obtain a mass have been universally agreed upon.

\section{An Ansatz for Enthalpy}

We are interested in finding the quantities of thermodynamic volume $V$ and
mass/enthalpy $M$ in the context of extended thermodynamic phase space \cite{Creighton1995,Caldarelli2000, Kastor2009} in which the
cosmological constant is understood as thermodynamic pressure: $P = -\Lambda/8 \pi G$.

It turns out we can make progress with a simple conjecture - that is, that the
scaling of mass is the same as that of asymptotically flat and AdS spacetimes, namely $L^{D-3}$, where $L$ is
some  fiducial length unit for the system.  Since
mass is a dimensionful quantity, for spherical black holes it can be expressed by combinations of two independent length scales: the event horizon radius $r_h$, and the cosmological length $l$ (where  $\Lambda \propto -1/l^2$).
We will make the natural assumption that mass is a function only of these two quantities, and point out where relevant what
happens if this assumption is dropped.
Entropy scales as $L^{D-2}$, since it is proportional to an area. In most of the solutions we consider, $S = A/4G$ where $A$ is the horizon area.
Theories with higher curvature terms will generate additional contributions to this expression but will not alter this scaling relation.

Putting the above requirements together and noting that $\Lambda$ scales as
$1/L^2$, we obtain via Euler's theorem \cite{Kastor2009}
 the Smarr formula
\begin{equation}
	(D-3) M = (D-2) T S - 2 PV
	\label{eqn:smarrgen}
\end{equation}
along with (in the absence of work terms) the first law
\begin{equation}
	dM = TdS + VdP
	\label{eqn:firstlaw}
\end{equation}
where $V$ is the thermodynamic volume conjugate to the pressure
\cite{Caldarelli2000, Kastor2009, Dolan2010}, and
\begin{equation}\label{liftemp}
	T = \left. \left( \frac{r}{l} \right)^{z + 1} \frac{1}{4 \pi} \sqrt{f'(r) g'(r)} \right|_{r = r_h}
\end{equation}
is the temperature. For spacetimes including Maxwell charges, the Smarr relation (\ref{eqn:smarrgen}) extends to
\begin{equation}
(D-3) M = (D-2) T S - 2 P V + (D-3) \Phi Q
\label{eqn:smarrgenwq}
\end{equation}
where  the Maxwell charge $Q$ is associated with the first thermodynamic law which is
\begin{equation}
d M = T dS + V dP + \Phi d Q.
\label{eqn:firstlawcharged}
\end{equation}

We pause to comment on an alternate Smarr relation that has been employed for Lifshitz spacetimes \cite{Bertoldi2009a,Dehghani2010,Liu2014b,Berglund2011,Dehghani2011c,Dehghani2013a,Way2012}, namely
\begin{equation}\label{eqn:othersmarr}
(D + z - 2) M = (D - 2) T S
\end{equation}
which is paired with a first law
\begin{equation}
d M = T d S.
\end{equation}
Here Euler's relation implies from (\ref{eqn:othersmarr}) that the mass term $M$ scales as  $L^{D+z-2}$, a relation inconsistent with the  $L^{D-3}$ scaling for $z=1$.  This is tantamount to assuming that the quantities in the Smarr relation depend on only the single
lengthscale  $r_h$, as  in asymptotically flat spacetimes. No $PV$ term arises as $l$ is not varied.
 
The distinction between the relations  (\ref{eqn:smarrgen}) and  (\ref{eqn:othersmarr}) is easiest to see by examining the Schwarzschild-AdS black hole. In this case we can define the metric for three horizon topologies (spherical, $k=1$; planar, $k=0$; hyperbolic, $k=-1$)
with
\begin{equation*}
f(r) =g(r) = 1 + k \frac{l^2}{r^2} - 2 \frac{m l^2}{r^{D-1}}
\end{equation*}
in (\ref{eqn:metric}), setting $z=1$. This yields a temperature
\begin{equation*}
T = \frac{1}{4\pi} \left[ (D-1) \frac{r_h}{l^2} + (D-3) k \frac{1}{r_h} \right]
\end{equation*}
and a mass
\begin{equation*}
M = (D-2) \frac{ \omega_{D-2} m }{16 \pi} = (D-2) \frac{ \omega_{D-2}}{16 \pi} \left( \frac{ r_h^{D-1}}{l^2} +  k r_h^{D-3} \right)
\end{equation*}
The entropy and pressure are
\begin{equation}\label{SPAdS}
S = \omega_{k,D-2} \frac{ r_h^{D-2}}{ 4}  \qquad P = -\frac{\Lambda}{8\pi G} = \frac{(D-1)(D-2)}{16 \pi l^2}
\end{equation}
and  geometric arguments \cite{Kastor2009} imply that the thermodynamic volume  coincides with the geometric volume
\begin{equation*}
V = \omega_{k,D-2} \frac{ r_h^{D-1}}{ (D-1)}
\end{equation*}
where $\omega_{k,D-2}$ is the surface area of the space orthogonal to fixed $(t,r)$ surfaces. Ignoring Eulerian scaling
we can write down a generalized Smarr relation
\begin{equation*}
M = a T S + b P V
\end{equation*}
where $a,b$ are undetermined coefficients.
We then have
\begin{align*}
 \left[(D-2) - a(D-1) - b(D-2)\right] & \frac{r_h^{D-1}}{l^2} \\
 + k \left[(D-2) -a(D-3)\right] & r_h^{D-3} = 0
\end{align*}
upon inserting the above relations.

We  see that for $k\neq 0$ the only solution to the above equation is $a = (D-2)/(D-3)$ and $b=-2/(D-3)$, consistent
with Eulerian scaling and the  Smarr relation (\ref{eqn:smarrgen}). However for planar ($k=0$) black holes, the second
relation is absent, yielding the one-parameter family of solutions
($a = (D-2)/(D-1) + \mathfrak{c}/(D-1)$, $b=-\mathfrak{c}/(D-2)$). In other words, there is a parameter's worth of ambiguity for   $k=0$. One way to resolve this is to set $\mathfrak{c}=0$, 
an approach commonly employed in Lifshitz spacetimes  for planar black holes
\cite{Bertoldi2009a,Dehghani2010,Liu2014b,Berglund2011,Dehghani2011c}.
Indeed, we see that (\ref{eqn:othersmarr}) is a special case of (\ref{eqn:smarrgen}) once it is recognized that $(D-1) PV = (D-2) TS$ for $z=1$ planar black holes. 

Since (\ref{eqn:othersmarr}) does not follow from either Eulerian scaling or
geometric considerations of the Komar formula \cite{Kastor2009} we regard 
(\ref{eqn:smarrgen}) as the appropriate Smarr relation for Lifshitz space times, valid for
all horizon topologies and asymptotics.  
This means choosing $\mathfrak{c}\neq 0$, which would appear to yield  a pressure that is ambiguous
(via $b$).

We can attempt to use the ideal gas law  to resolve this ambiguity.
For asymptotically uncharged planar or toroidal AdS black holes, the general form of the equation of state   is equivalent to that of an ideal gas $Pv =T$ \cite{Altamirano2014}, recognizing that the specific
 volume $v = V/N$.  The number of degrees of freedom $N$ is identified
with the number of degrees of freedom associated with the black hole, which is proportional to its horizon area in Planck units; hence $N\propto S$ and we obtain  $PV \sim TS$ for planar/toroidal black holes.
Since $T$ can be unambiguously computed from the metric, we will obtain the proportionality constant in this relation provided the conserved charges (particularly the mass) are independently known.
For Lifshitz black holes this is not in general the case. 
However, if we begin with the equation (\ref{eqn:smarrgenwq}), then the statement that for $k=0$ the ideal gas law $PV \sim NT$ holds is equivalent to fixing the factor in $N \propto S$ through equation (\ref{eqn:othersmarr}).
We will be fixing the ideal gas law to be $2(D+z-2)PV=(D-2)(z+1)TS$ throughout, pausing to compare the results of our approach with (\ref{eqn:othersmarr}) where appropriate.

For the Smarr equation (\ref{eqn:smarrgen}) and the $1^{st}$ law, we will determine the thermodynamic mass and volume
given temperature, entropy, and pressure.  Provided that the black hole has two independent length scales, one from the  event horizon radius $r_h$ the other from the cosmological constant $\Lambda \propto -1/l^2$,  we can avoid multiple solutions for the mass.  However in some (perhaps unusual) exact solutions that we examine below, the black hole event horizon radius
and AdS length scale are no longer independent parameters.
In this case the Smarr relation is degenerate
with the first law, and we can only specify a mass as dependent on the thermodynamic
volume (or vice versa).

\subsection{Basic Method}

Consider the case where the entropy depends only on the horizon length scale  and pressure on the cosmological length scale. Motivated by (\ref{liftemp}) we will  assume that the temperature can be expressed as  $T = c_0 r_h^{\tilde{\beta}} l^{\tilde{\alpha}}$ where $\tilde{\beta},\tilde{\alpha}$ are real parameters;  this assumption holds for at least planar black holes.

To illustrate the method, we shall use the aforementioned equations and assumptions to construct two additional formulae
which will force solutions to obey the first law of thermodynamics. Regarding $M$ and $V$ as unknown quantities,
from (\ref{eqn:firstlaw}) we obtain two equations
\begin{equation}
	\frac{\partial M}{\partial l}  =  V \frac{\partial P}{\partial l}  \qquad
	\frac{\partial M}{\partial r_h} = T \frac{\partial S}{\partial r_h}
	\label{firstlaweqs}
\end{equation}
provided $l$ and $r_h$ are independent. Employing the ansatz $M = M_0 r_h^{\beta} l^{\alpha}$ yields
\begin{align}
	M_0 &= \frac{V}{\alpha l^{\alpha - 1} \cdot r_h^{\beta}} \frac{\partial P}{\partial l} \nonumber\\
	V &= \frac{T \cdot \alpha \cdot r_h}{\beta \cdot l} \cdot \frac{\partial S}{\partial r_h} \left( \frac{\partial P}{\partial l}\right)^{-1}
	\label{mandvsol}
\end{align}
where $M_0$ is a dimensionless constant independent of $r_h$ and $l$. Note that on dimensional grounds $\alpha + \beta = D - 3$.

These formulae considerably simplify whenever $\Lambda \propto 1/l^2$ and $S \propto r^{D-2}_h$, since in this situation
\begin{equation}
\label{eqn:easyassumptions}
r_h \frac{\partial S}{\partial r_h} = (D-2) S  \qquad     \frac{\partial P}{\partial l} =  - \frac{2 P}{l}
\end{equation}
Hence in general from (\ref{mandvsol}) we have
$$
V = \frac{T \cdot \alpha \cdot r_h}{\beta \cdot l} \cdot \frac{\partial S}{\partial r_h} \left( \frac{\partial P}{\partial l}\right)^{-1}
= -\frac{\alpha(D-2)}{\beta} \frac{TS}{2P}
$$
yielding a formula for the thermodynamic volume in terms of $(T,S,P)$,  generalizing the relation noted above for AdS planar black holes.  The
 Smarr formula (\ref{eqn:smarrgen})  then gives
$$
M = \frac{D-2}{D-3} TS  - \frac{2}{D-3} PV
= \frac{D-2}{D-3} TS\left[1+ \frac{\alpha}{\beta} \right] = \frac{D-2}{\beta} TS
$$
since $\alpha + \beta = D-3$.
Assuming the scaling property
\begin{equation}
\label{eqn:easytempcondition}
	[T] = r_h^{z}/l^{z+1}
\end{equation}
based on the expression (\ref{liftemp}), we obtain
\begin{equation}
	[M] = r_h^{z+D-2}/l^{z-1-1+3} = r_h^{z+D-2}/l^{z+1}
	\label{eqn:mscaling}
\end{equation}
and so $\beta= D+z-2$.  This gives
\begin{equation}
\label{eqn:masseasy}
M =  \frac{D-2}{D+z-2} TS
\end{equation}
which is the same as (\ref{eqn:othersmarr}), as well as
\begin{equation}
\label{eqn:voleasy}
V = \frac{(z+1)(D-2)}{D+z-2} \frac{TS}{2P}
\end{equation}
for the volume.

The preceding relations are sufficient to solve for $(M,V)$ since for a given Lifshitz black hole, the quantities
$(T,S)$  are both relatively easy to compute, and the relationship between  $\Lambda$ and $l$ is known; note that although
$\Lambda \propto 1/l^2$, the coefficient in general is not  the same as in the AdS case (\ref{SPAdS}). Note also that when $\beta = 0$ the system of equations (\ref{mandvsol}) no longer yields a solution.  This
corresponds to the case in which mass does not vanish as the horizon radius goes to zero. We shall not consider
such cases in this paper in detail, as they imply that the mass $M$ is no longer solely that of the black hole.
This is not meant to suggest that this method will not apply for these solutions, which may be an interesting topic for future work;
a mass for soliton solutions such as those described in \cite{Basu2010,Dias2011a} can be obtained via our method when $T S = 0$
and we expect that the same approach could be applied to find a mass for the numerical Lifshitz solitons in \cite{Mann2011b}.
Generally these solutions will reduce to the two-lengthscale approach elaborated upon below.

We also note that higher curvature terms in the action will spoil this simplification since
entropy will contain multiple terms with different scaling in $r_h, l$. In addition,
the temperature scaling requirement is very strict; it effectively restricts us to
metrics of the form $f(r) = 1 + m (l/r)^{\mathfrak{p}}$, which yield a $T$ proportional
to $r^z/l^{z+1}$ for any $\mathfrak{p}$. Adding a term, say, $k l^2/r^2$, will modify
this temperature relationship.

\subsection{A More General Approach}

Our ansatz can be improved to deal with more complicated black holes by
expressing the temperature, mass, and volume in a series of powers of
horizon size and AdS length.  We write
\begin{equation*}
	T = \sum_i T^{(i)} = \sum_i T_0^{(i)} r_h^{\tilde{\beta}_i} l^{\tilde{\alpha}_i}
\end{equation*}
and
\begin{equation*}
M = \sum_i M_0^{(i)} r_h^{{\beta}_i} l^{{\alpha}_i}
\qquad
V = \sum_i V^{(i)} r_h^{\hat{\beta}_i} l^{\hat{\alpha}_i}
\end{equation*}
where $T_0^{(i)}$, $M_0^{(i)}$ and $V^{(i)}$ are all dimensionless coefficients whose
details depend on the black hole under consideration. Requiring the Smarr relation (\ref{eqn:smarrgen}) and first law
to hold  then  implies the relations
\begin{equation*}
\alpha_i = \tilde{\alpha}_i = \hat{\alpha}_i + 2
\qquad
\beta_i = \hat{\beta}_i = \tilde{\beta}_i + \left( D - 2 \right)
\end{equation*}
and
\begin{align}
	M_0^{(i)} &= \frac{V^{(i)}}{\alpha_i l^{\alpha_i - 1} \cdot r_h^{\beta_i}} \frac{\partial P}{\partial l} \\
	V^{(i)} &= \frac{T_0^{(i)} \cdot \alpha_i \cdot r_h}{\beta_i \cdot l} \cdot \frac{\partial S}{\partial r_h} \left( \frac{\partial P}{\partial l}\right)^{-1}
	\label{mandvsol2}
\end{align}
provided $S$ depends only on $r_h$,  not $l$.

Further generalizing to the case where the entropy is also
a sum of terms proportional to $r_h$ and $l$ to some powers, we have
\begin{equation}
	\label{eqn:fullseriesexp}
	TS =\sum_i T^{(i)}  \sum_j S^{(j)} = \sum_{i j} T_0^{(i)} S_0^{(j)} r_h^{\beta_{i j}} l^{\alpha_{i j}}
\end{equation}
with the coefficients $\beta_{i j}$ and $\alpha_{i j}$ corresponding to the powers of $r_h$ and $l$ that appear in this expansion.

Writing the Smarr relation (\ref{eqn:smarrgen})  as
\begin{align}\label{alpbetij}
	(D-3) M &= (D-3)  \sum_{i,j} M_0^{(ij)} r^{\beta_{i j}} l^{\alpha_{i j}} \\
	&= (D-2) \sum_i
	\sum_j T^{(i)} S^{(j)} - 2 P \sum_{i,j} V^{(ij)} \nonumber
\end{align}
yields the solution
\begin{align}
	\label{mandvsol3}
	M_0^{(ij)} &= \frac{T^{(i)}}{\beta_{ij} l^{\alpha_{ij}} \cdot r_h^{\beta_{ij}-1}} \frac{\partial S^{(j)}}{\partial r_h} \\
	V^{(ij)} &= \frac{T^{(i)}}{2 P \beta_{ij} } \bigg[ (D-2) \beta_{ij} S^{(j)}  \nonumber \\
	&  \quad \quad \quad \quad -  (D-3) r_h \frac{\partial S^{(j)}}{\partial r_h}  \bigg] \nonumber
\end{align}
upon matching coefficients.

If the series above are infinite (which may prove applicable for approximating numerical black holes), we see that
if the small-$r_h$ series for $T$ and $\partial S / \partial r_h$ are both convergent, and one or more
of them are absolutely convergent, then the series solution for mass in terms of  $r_h$ will also be convergent.
A near-horizon expansion could then be performed for numerical solutions, yielding a convergent
quantity for the mass without needing to obtain the temperature in exact closed form. We shall not consider numerical solutions in this paper.

Note that the form for $M$ in equation (\ref{mandvsol3})
is equivalent to that of approaching the problem in the context where the
thermodynamic variables are assumed independent, and where
\begin{equation*}
\left( \frac{\partial M }{\partial S} \right)_{V} = T
\end{equation*}
is integrated over $r_h$, in the case where the lengthscale $r_h$ is independent of $l$.
 In the next subsection we consider
the case where it is not,  and
provide explicit examples of such cases throughout the paper, starting in
section \ref{mannsol}.

We close this subsection with  some remarks on the conditions for the positivity of the mass and volume
using this approach. First, we only consider $\Lambda < 0$, so that  pressure
is a positive quantity and therefore the derivative of pressure with respect to $l$
is negative. Second, the derivative of the entropy with respect
to the horizon radius is also assumed positive.  This reasonable assumption follows from the
microstate counting argument for entropy, which suggests that a larger black hole should have higher
degeneracy and therefore greater entropy. Third, we assume a positive temperature
for the black hole.  Finally, we assume that the various thermodynamic quantities have a well-defined
limit as $\Lambda \to 0$ (or $l \to\infty$), implying that $\alpha_i < 0$.

With these constraints we find that mass is always positive when $TS$ is dependent on horizon radius to a positive power.
We do not yet have suitable physical conditions on positivity for the volume.

\subsection{Dependent Length-scales}
\label{dependentlength}

We must also address the difficulty that appears when the length-scales $r_h$ and $l$ are dependent.
This case occurs for a number of exact black hole solutions in
asymptotically Lifshitz spacetimes.
In this circumstance we know that the mass must scale like $l^{(D-3)}$ and the entropy must
scale like $l^{(D-2)}$ since there are no other length-scales, so by taking the derivative
with respect to $l$, the first law (\ref{eqn:firstlaw}) reduces to
\begin{equation}
\label{eqn:dependent}
(D-3) M/l = (D-2) T S/l - 2 V P/l
\end{equation}
to coincide with the Smarr relation  (\ref{eqn:smarrgen}). The solution in this case
has $M = M(V)$, while $V = V(l)$, so we obtain a one-parameter family of
valid solutions in this underconstrained system. In this situation the thermodynamic method fails to independently define a mass, and an alternate approach must be found.

One approach is to introduce a fictitious
parameter $\tilde{m}$ in the metric function that temporarily separates the two length-scales. This allows us to use
the thermodynamic approach to obtain a mass and volume, after which
 we take the limit $\tilde{m} \to 0$.
This is still hardly unique, but we can attempt to justify our choice
of ``mass parameter'' by using notions of how mass conventionally scales.
Typically we see that it appears in the metric function as a term $M/r^{D-1} = \tilde{m} (l/r)^{(D-1)}$
where $\tilde{m}$ is dimensionless. For anisotropic spacetimes the form of
the mass term is an open question but as we will see below, a plausible
ansatz is $\tilde{m} (l/r)^{(D + z -2)}$.
Note that though this approach appears to alter the temperature of the black hole,
it only alters the scaling of the temperature; upon substituting $\tilde{m}=0$ in the
final temperature, all fictitious mass methods agree on the temperature,
entropy, and pressure of the black hole system.

 Another way to resolve the ambiguity of obtaining a thermodynamic mass ($M$) and a volume ($V$) is to regard $r_h$ as   independent from $l$ in the first law, obtain $M$ and $V$, and then take the limit that yields $r_h$ as a function of $l$ for
 the black hole solution to hold.  Specifically, we   integrate  $dM$ with respect to $r_h$ from the first law 
 \begin{align}
M(r_h, l) &= \int d {r_h} \frac{\partial M(r_h, l)}{\partial r_h} \nonumber \\
	  &= \int  d {r_h}  T(r_h,l) \frac{\partial S(r_h)}{\partial r_h}
 \label{thermoM}
\end{align}
 to obtain $M$, holding $l$ fixed. We then vary $M$ with respect to $l$ 
 \begin{equation}
\frac{\partial M(r_h,l)}{\partial l} = V(r_h,l) \frac{\partial P(l)}{\partial l},
 \label{thermoV}
\end{equation}
 to obtain $V$. The last step is to  substitute the function $r_h = r_h(l)$ that is consistent with the black hole solution.

A third approach simply makes a firm choice of a thermodynamic parameter to resolve the ambiguity.
If we suppose that each independent parameter
added to the action that generates a lengthscale also generates a thermodynamic quantity
appearing in the Smarr relation (along with its conjugate potential), we
can make a choice to eliminate certain Smarr terms for solutions with
fewer independent lengthscales than is usual. 
For example, if the Ricci scalar generates the free lengthscale $r_h$ and entropy $S$ for a black hole spacetime,
adding a cosmological constant yields a new lengthscale $l$, as well as a pressure $P$.
A Maxwell field will yield a lengthscale $q$ which corresponds to the thermodynamic charge $Q$.
Higher curvature terms have similarly been conjectured to generate terms that appear in the Smarr relation.

The most compelling choices will be to firmly fix $M=0$ or $V=0$ which we consider in some examples below; 
in these cases $dM = 0$ and $dV = 0$ respectively,
and the remaining thermodynamic quantities can then be computed via our protocol.

We note that the freedom in choosing the scaling of the $\tilde{m}$
term in the fictitious mass approach is equivalent to fixing one of
$M$ or $V$ in terms of the other; this approach is mentioned only
because it may provide some conceptual insight.

\subsection{Charge}
\label{chargesoln}

Charge will add a $\Phi Q$ term
to the Smarr relation as well a $\Phi dQ$ term to the first law,
 where $\Phi$ is the value of the electromagnetic potential at the horizon.
To obtain the charged Smarr formula, we
first apply the above algorithm when $Q=0$ to obtain a solution which is uncharged.
We assume that  charge is an independent thermodynamic
quantity from $r_h$ and $l$; a reasonable assumption given that the horizon condition $f(r) = 0$
fixes at most one length scale in terms of the others.

We then solve  the Smarr relation (\ref{eqn:smarrgenwq})
\begin{equation}
\label{Smarrchaged}
(D-3) M = (D-2) T S - 2 P V + (D-3) \Phi Q
\end{equation}
along with equation (\ref{firstlaweqs}) and with the additional  relation
\begin{equation}
\frac{\partial M}{\partial Q} dQ = \Phi dQ \label{pMovpQ}
\end{equation}
for
$\Phi Q$. This can be done in a similar manner as before (expanding $M$
in a series depending on charge, then eliminating the linear coefficients
and using consistency of the powers in the sums to obtain a unique solution).
We do not find enough freedom to obtain $\Phi$ separately from $Q$,
but we are able to find the product's respective dependence on $l$ and $r_h$, so we
obtain $\mathfrak{k} \Phi$ and $Q/\mathfrak{k}$ up to some constant coefficient $\mathfrak{k}$.

The precise method involves
splitting the $M$ and $TS$ terms into $Q$-dependent parts and $Q$-independent parts.
Since we have assumed $S$ is independent of $Q$, we take $T^{(Q)} \equiv T - T^{(q=0)}$
where $Q$ is the Maxwell charge;
from the scaling of the Maxwell equations it behaves as a third lengthscale
$L^{D-3}$. At this stage we know only the scaling of $Q$, so we can identify $Q \sim q$ where
the free parameter in the metric function, $q$  has $L^{D-3}$ scaling.

Denoting the power with which $q_i$ ($i$ indexes multiple charges) appears in $T^{(Q_i)}$ by $\mathfrak{a}$,
we can perform a power series in $q_i$,
{\begin{align}
&M = M_0 + \sum_{i = 1}^N M_i(r_{h},l) q^{\mathfrak{a}_i}_i, \; M_{i} = \sum_{j,k} M_{i}^{(jk)} r_{h}^{\gamma_{jk}} \; l^{\delta_{jk}},\\
&T = T_0 + \sum_{i = 1}^N T_i(r_{h},l) q^{\mathfrak{a}_i}_i, \; T_{i} S = \sum_{jk}T_{i}^{(j)} S_{0}^{(k)} \; r_{h}^{\gamma_{jk}} \; l^{\delta_{jk}}  \\ 
&V = V_0 + \sum_{i = 1}^N V_i(r_{h},l) q^{\mathfrak{a}_i}_{i} , V_{i} P = \sum_{jk} V_{i}^{(j)} P_{0}^{(k)} \; r_{h}^{\gamma_{jk}} \; l^{\delta_{jk}} \label{temp1}\\
&\Phi = \sum_{i = 1}^N \Phi_i(r_{h},l) q^{\mathfrak{a}_i-1}_{i}, \; \; \Phi_{i} = \sum \Phi_{i}^{(jk)} \; r_{h}^{\gamma_{jk}} \; l^{\delta_{jk}}
\end{align}}
where $M_0$, $V_0$ and $T_0$ satisfy the uncharged Smarr relation. {Since $[q_i^{\mathfrak{a}_i}] = \mathfrak{a}_i (D-3), \gamma_{jk} + \delta_{jk} = - (\mathfrak{a}_i -1)(D-3)$.}
The Smarr relation and $1^{st}$-law
then yield for $i\ne0$
\begin{align}
(D-3) M_i &= (D-2) T_i S - 2 V_i P + (D-3) \frac{\Phi_i Q_i}{ q_i } \label{temp4} \\
\frac{\mathfrak{a}_i M_i}{Q_i} &= \frac{\Phi_i }{ q_i} . \label{temp3}
\end{align}
The latter of the two equations allows us to eliminate $\Phi_i Q_i$ from the former,
and we can use our previous technique to solve for the mass under the modified
Smarr relation
\begin{equation}
(D-3)(1 - \mathfrak{a}_i) M_i = (D-2) T_i S - 2 V_i P .
\end{equation}
The electric potential $\Phi_i$ can then be determined from equation
(\ref{temp3}) and a Maxwell charge relation
\begin{equation}
Q = \frac{1}{4 \pi} \int *F.
\end{equation}
(or something similar, in the case where the gauge field is coupled
to other fields such as a dilaton).
For example, when $l$ and $r_h$ are independent, we need only use
{\begin{equation}
	\frac{\delta_{j k} M_{i}}{ l } = -2 \frac{ V_i P}{ l } \label{temp2}
\end{equation}}
where {$\delta_{j k}$} is the power of $l$ in the $T^{(j)}_i S^{(k)}$ term
[after expanding $M_i$ and others into series in $l, r_h$ similar to equation
(\ref{eqn:fullseriesexp})], to obtain
\begin{align}
&\Phi_i Q_i = \mathfrak{a}_i M_i q_i \nonumber \\
&M_i = \frac{(D-2)}{(D-3)(1 - \mathfrak{a}_i) - {\delta_{jk}}} T_i S \label{eqn:phiQresult} \\
&V_i = - \frac{ \delta_{j k} M_i}{2 P} \nonumber
\end{align}
We consider a Maxwell charge which yields $\mathfrak{a} _{i}= 2$.
In this case, we can simplify even further under the assumptions
(\ref{eqn:easyassumptions}) and (\ref{eqn:easytempcondition}) to obtain
\begin{equation}
	\label{eqn:simplifiedSmarr}
M= \frac{D-2}{D+z-2} TS + \sum_{i = 1}^N  \frac{2D + z - 5 + \delta_{jk}}{2 (D + z - 2) } \Phi_{i} Q_{i}
\end{equation}
where {$Q$} is the Maxwell charge appearing in the first thermodynamic law
\begin{equation}
	d M = T d S +   \sum_{i = 1}^N \Phi_i  dQ_i
\label{eqn:fstlawwthpv}
\end{equation}

\section{Exact Solutions}

Here we apply our approach to solve for the mass and volume for a variety of
exact solutions.  To test our method, we begin with the well-understood Reissner-Nordstr\"om AdS black hole,
and the more recently studied AdS-Taub-NUT (Newman, Unti, Tamburino) case \cite{Johnson2014}.  We then
move on to several different examples of exact Lifshitz symmetric black hole solutions \cite{Mann2009,Bertoldi2009,Gim2014,Pang2009}.
In a number of these theories, the derivation of the mass has been a subject of controversy
where differing proposed methods yield different masses, and there is little consensus on which of these masses is the most physically meaningful.
For example, the masses presented in the higher curvature Lifshitz theory from \cite{Gim2014} 
use a method \cite{Hohm2010} which yields a nonunique counterterm and therefore a nonunique mass.
For this theory we consider exact solutions for both $z = 3$, $D = 3$ and  $z=2$, $D = 5$. We can attempt to use the
entropy, cosmological constant, and temperature to obtain a mass and compare with the masses obtained by other
methods (e.g. via a quasilocal formalism).

We will also check the conjectured reverse isoperimetric inequality \cite{Cvetic2010} $\mathcal{R} \ge 1$ for each of these solutions, where
\be\label{eq:ipe-ratio}
\mathcal{R} = \left(\frac{(D-1)\mathcal{V}}{\omega_{k,D-2}}\right)^{\frac{1}{D-1}}\left(\frac{\omega_{k, D-2}}{\mathcal{A}}\right)^{\frac{1}{D-2}}
\ee
with
\be
\omega_{1,D} = \frac{2\pi^{\frac{D+1}{2}}}{\Gamma\left(\frac{D+1}{2}\right)}
\ee
where  $\mathcal{A}$ is the horizon area.  This inequality is essentially a statement of the amount of entropy a given
black hole can contain. If the ratio is greater than one, the conjecture implies that
the maximal amount of entropy for that volume has not yet been reached. When
the ratio is unity, the statement is that the given solution has reached the
maximal amount of entropy. In its original form, the Schwarzschild-AdS black
hole was seen to maximize the ratio ($\mathcal{R} = 1$), while Kerr-AdS black
holes with finite rotation all exhibited $\mathcal{R} >1$. A new class of super-entropic black hole solutions
was recently obtained for which $\mathcal{R} <1$ by taking a new ultraspinning limit of the Kerr-AdS solution in
$D$-dimensions \cite{Hennigar2014}.

\subsection{z=1, D=4}

The Reissner-Nordstr\"om AdS black hole is the $z=1$, $D=4$,
metric from equation (\ref{eqn:metric}) with $k=1$ where
\begin{equation}
f(r) = 1 + k \frac{l^2}{r^2} - \frac{2 m l^2}{r^3} + \frac{q^2 l^2}{r^4}
\label{eqn:fr}
\end{equation}
which has the solution for the mass parameter (taking $q=0$ and $f(r_h) = 0$) of $m = \frac{r_h^3}{2 l^2} +
\frac{r_h}{2}$. Note that $m$ in this case coincides with the ADM mass of the black hole; $m
= {\cal M}$.
In addition, $\Lambda = -3/l^2$.

We now show that the thermodynamic method we have illustrated above
will independently yield the correct ADM mass using the pressure,
temperature, and entropy of this black hole.

The pressure is $P = 3/8 \pi l^2$,
while $f(r)$ given in (\ref{eqn:fr})
yields a temperature of
\begin{equation*}
T = \frac{3 r_h}{4 \pi l^2} + \frac{k}{4 \pi r_h}.
\end{equation*}
Finally, the entropy of this black hole is
$A/4 = \omega_{k,2} r_h^2 / 4$,
 where $\omega_{k,2}$ is the surface area of the
2-surface for the $k$ topology.
 Since the temperature expands into two terms, we
apply equation (\ref{mandvsol3}) where $i$ ranges
from 0 to 1 while $j$ only takes a single value of 0.
The coefficients are then $\alpha_{00} = -2, \beta_{00} = 3$
and $\alpha_{10} = 0, \beta_{10} = 1$ yielding
\begin{equation*}
M_0^{(00)} = \frac{\omega_{k,2}}{8 \pi} \qquad M_0^{(10)} = \frac{k \omega_{k,2}}{8 \pi}
\end{equation*}
and
\begin{equation*}
V_0^{(00)} = \frac{\omega_{k,2}}{3} \qquad V_0^{(10)} = 0
\end{equation*}

We therefore obtain a mass and volume of
\begin{equation}
	M =  \frac{ \omega_{k,2} r_h^3}{8 \pi l^2} + \frac{k \omega_{k,2} r_h}{8 \pi} \qquad V = \frac{\omega_{k,2} r_h^3}{3}
\end{equation}\label{RNMVol}
for $Q=0$. The conclusion here is that our
thermodynamic method agrees with the ADM mass of this black hole.
It is straightforward to show that $\mathcal{R} = 1$ when $Q=0$,
so the reverse isoperimetric inequality is saturated for the Schwarzschild-AdS black hole.

We then apply the remainder of our thermodynamic approach when $Q \ne 0$; see
section (\ref{chargesoln}).
The charge term in the temperature is now
\begin{equation*}
T_1 = - \frac{ {q^2}}{4 \pi r_h^3}
\end{equation*}
and $\alpha_1 = 0$,  and the Maxwell charge is
\begin{equation}
Q = \frac{\omega_{k,2}}{4 \pi}q. \label{IIIAQ}
\end{equation}
From (\ref{eqn:phiQresult})
we find that
\begin{equation}
M = \frac{\omega_{k,2} r_h^3}{8 \pi l^2} + \frac{k \omega_{k,2} r_h}{8 \pi} + \frac{\omega_{k,2}  {q^2}}{ 8 \pi r_h} \label{IIIAMwtq}, \; \; \; \; \; V = \frac{\omega_{k,2} r_{h}^{3}}{3}
\end{equation}
and by inserting (\ref{IIIAQ}) and (\ref{IIIAMwtq}) into (\ref{pMovpQ}) we
obtain an electric potential at the horizon of
\begin{equation}
\Phi = \frac{q}{r_h} \label{IIIAphi}
\end{equation}
which agrees with the potential calculated from the Maxwell equation. 
Note that the thermodynamic volume of equation (\ref{IIIAMwtq})
is the same as that in the uncharged case.

Note that in the $k=0$  planar case, we can apply the simplified Smarr relation
 (\ref{eqn:simplifiedSmarr}); specifically, (\ref{IIIAMwtq}) reduces to
\begin{equation*}
M = \frac{2}{3} TS + \frac{2}{3} \Phi Q
\end{equation*}

\subsection{z=1, D=3}

 The non-rotating BTZ black hole \cite{Banados1992} provides another obvious check on the consistency of our method.
 The metric function is \cite{Carlip2005}:
 \begin{equation*}
 f(r) = 1 - \frac{8 M l^2}{r^2}
 \end{equation*}
 and $P = 1 / 8 \pi l^2$, $S = \pi r_h / 2$, and $T = r_h / 2 \pi l^2$.

Inserting these values of pressure, entropy,
 and temperature into equation (\ref{mandvsol3}) yields
 $M_0^{(00)} = 1/8$ and $V_0^{(00)} = \pi$.
 Upon using equation (\ref{alpbetij}), our procedure returns
 \begin{equation}
 \label{eqn:btzth}
 M = \frac{ r_h^2}{ 8 l^2} \qquad V = \pi r_h^2
 \end{equation}
and we see that the parameter $M$ agrees with the ADM mass $\mathcal{M}$
 and the Brown-York Mass \cite{Carlip1995}.  In addition, the thermodynamic volume is the area of a disc of radius $r_h$.
 The isoperimetric inequality of equation (\ref{eq:ipe-ratio}) is similarly saturated for this black hole;
 we obtain $\mathcal{R} = 1$.

\subsection{z=1, D=4}

Here we look at the AdS-Taub-NUT
spacetime, in the form presented in \cite{Chamblin1999}.
 Recently, the extended pressure-volume first law was
used to identify a thermodynamic volume for these spacetimes, given a mass, temperature,
entropy, and pressure \cite{Johnson2014}.   This case is somewhat less trivial than the preceding examples
since the entropy now depends on two length scales.
We therefore seek to show that our methods can replicate this result.

The metric in Euclidean form is
\begin{align*}
ds^2 = F(r) & \left( d\tau + 2n \cos(\theta) d\phi \right)^2 + \frac{dr^2}{F(r)} \\
& + (r^2 - n^2) \left( d\theta^2 + \sin^2(\theta) d\phi^2 \right)
\end{align*}
where
\begin{align*}
F(r) \equiv \frac{ (r^2 + n^2) - 2 m r + l^{-2} (r^4 - 6 n^2 r^2 - 3n^4)}{r^2 - n^2}
\end{align*}
The relevant quantities are
\begin{equation}\label{TSnut}
T = \frac{1}{8\pi n}, \quad S = 4\pi n^2 \left[ 1 - \frac{6 n^2}{l^2} \right]
\end{equation}
and $P = 3/8 \pi l^2$.
In the Taub-NUT solution $r _h= n$, implying the $S^2$ is of zero size.
We therefore consider two length scales, $l$ and $n$;  the mass is a function of both.

 We see from (\ref{TSnut})
that the assumption (\ref{eqn:easyassumptions}) for entropy  does not hold.
Our method thus requires the more general equations (\ref{mandvsol3}).
Because $r_h = n$ we replace all instances of
$r_h$ in equation (\ref{mandvsol3}) with $n$.
Then, given the temperature, entropy, and
pressure, we obtain a mass upon substitution into
equation (\ref{mandvsol3}). Putting the resultant $M_0^{(00)}$
and $M_0^{(10)}$ into equation (\ref{alpbetij}) yields
$M = n - 4n^3/l^2$ (equal to the
mass parameter in the AdS-Taub-NUT) as well as a volume $V =
-8\pi n^3/3$, in agreement with \cite{Johnson2014}.

The negative value for the volume has been interpreted in terms of the work the universe does to create
the black hole \cite{Johnson2014}.  Recall that enthalpy is the sum of the energy required to both make
a system and to place it in an environment.  In the AdS-Schwarzschild scenario, a volume of
radius $r_h$ is excised from  empty AdS to produce the black hole, requiring work to be done on AdS.  However
in  AdS-Taub-NUT the thermodynamic scenario involves adding the  volume $V =
-8\pi n^3/3$ to  empty (Euclidean) AdS space to create a
Taub-NUT black hole metric, since the $r_h=n$ is of zero size; this means that work is done by AdS to create the
black hole \cite{Johnson2014}.

We see that in this scenario we obtain the agreed-upon mass and volume
via the thermodynamic method for the AdS-Taub-NUT spacetime.

\subsection{z=2, D=5}

A certain class of Lifshitz black holes are exact solutions to the field equations
that follow from the action \cite{Gim2014}
\begin{align}
	\mathcal{I} = \int d^D x & \sqrt{-g} \left( \frac{1}{\kappa} \left[R - 2 \Lambda \right] + a R^2 + b R^{\mu \nu} R_{\mu \nu} \right. \nonumber \\
	& \quad \left. + c \left[ R_{\mu \nu \rho \sigma} R^{\mu \nu \rho \sigma} -4 R_{\mu \nu} R^{\mu \nu} + R^2 \right] \right)
	\label{eqn:GimAction}
\end{align}
and have the form
\begin{equation*}
ds^2 = -\left(\frac{r}{l}\right)^{2z}f(r) dt^2 + \frac{l^2 dr^2}{ g(r) r^2} + d\vec{x}^2
\end{equation*}
where
\begin{align}
\label{eqn:lifd5hc}
f(r) = g(r) = \left( 1 - \frac{ m l^{5/2}}{r^{5/2}} \right)
\end{align}
with $a = -16 l^2/725, b = 1584 l^2/13775, c = 2211l^2/11020$.
The general class of exact black hole solutions that exist by tuning $z$ and
the higher curvature parameters was originally presented in \cite{Ayon-Beato2010}.

Here we will examine only the specific case of equation (\ref{eqn:lifd5hc})
as presented in \cite{Gim2014}. The authors of that paper find
\begin{equation*}
	S = \frac{396 r_h^3 \pi  {\omega_{0,3}}}{551}
\end{equation*}
for the entropy,  where $\omega_{0,3}$ is the surface area of the constant $(t,r)$ toroidal section.
 The temperature and cosmological constant are
\begin{equation*}
	T = \frac{ 5 r_h^2}{ 8 \pi l^3},  \quad %
	\Lambda = \frac{-2197}{551 l^2}
\end{equation*}
The pressure is therefore
\begin{equation*}
P = \frac{2197}{4408 \pi l^2}.
\end{equation*}

Because these planar black hole solutions satisfy the assumptions for the temperature
and entropy from equations (\ref{eqn:easyassumptions}) and (\ref{eqn:easytempcondition}),
the  reduced Smarr relation (\ref{eqn:othersmarr}) holds, and can be used 
 to yield mass from only $T$ and $S$.
Therefore, substituting into equations
(\ref{eqn:masseasy}) and (\ref{eqn:voleasy}) gives
\begin{align}
	V &= \left( \frac{5 r_h^2}{8 \pi l^3}   \frac{1188 \pi r_h^2
	\omega_{0,3}}{551} \right)   r_h (-3)  \left( \frac{5}{\pi l^2}
	\frac{-2197}{2204} \right)^{-1} \nonumber \\
	&= \frac{1782}{2197} \cdot \frac{r_h^5
	\pi \omega_{0,3}}{l}
\end{align}
and
\begin{equation}\label{Gim-mass-d5}
	M = \frac{1782}{2197}   \frac{r_h^5 \pi \omega_{0,3}}{l} \cdot \left(
	\frac{2197 l \omega_{0,3}}{3\times 2204 \pi l^2 } \right) =
	\frac{297}{1102} \cdot \frac{r_h^5 \omega_{0,3}}{l^3}
\end{equation}
which is precisely the mass obtained in \cite{Gim2014}.  The thermodynamic mass and volume, obtained from the reduced Smarr formula (\ref{eqn:othersmarr}), also satisfy the general Smarr formula (\ref{eqn:smarrgen}) and the thermodynamic first law (\ref{eqn:firstlaw}).

Although it  has been stated  \cite{Hohm2010} that (\ref{Gim-mass-d5})  agrees with the Brown-York mass for this black hole,
appropriate counterterms are not yet
known for this spacetime.
Rather, the required counterterms were determined by demanding the first law to be satisfied  \cite{Hohm2010},  so the mass
in \cite{Gim2014} is not independent of our thermodynamic considerations.
Nonetheless, it is compelling that we have found a mass, independent of
any quasilocal formalism, that agrees with the method of \cite{Gim2014}.

The reverse isoperimetric inequality $\mathcal{R} \ge 1$ is violated in this case. We obtain
\begin{equation*}
	\mathcal{R} = 3 \left( \frac{11 \pi \cdot 2^{3}}{13^{3}} \frac{r_h}{l} \right)^{\frac{1}{4}}
\end{equation*}
and so for sufficiently small $r_h$ we will have $\mathcal{R}  < 1$.

\subsection{z=3, D=3}
We can examine the $z=3$, $D=3$ higher curvature Lifshitz solution which was 
found in New Massive Gravity \cite{Ayon-Beato2009}: 
\begin{equation*}
ds^2 = -\left(\frac{r}{l}\right)^{2z}f(r) dt^2 + \frac{l^2 dr^2}{r^2  g(r) } + r^2 d\phi^2
\end{equation*}
with
\begin{align*}
	f(r) = g(r) = \left( 1 - m\frac{l^2}{r^2} \right)
\end{align*}
where the action is the same as in equation
(\ref{eqn:GimAction}), but the parameters
$a = -3 l^2/4 \kappa, b = 2 l^2/\kappa$, and $c = 0$, while the cosmological
constant is $\Lambda = -13/2 l^2$.
The thermodynamic parameters  of this exact solution are 
$$
S= 2\pi r_h \qquad P = 13/16 \pi l^2 \qquad T = r_h^3 / 2 \pi l^4.
$$

We satisfy the assumptions from equations (\ref{eqn:easyassumptions}) and (\ref{eqn:easytempcondition}), so the Smarr relation again reduces to the simpler form
(\ref{eqn:othersmarr}). Substituting $T$ and $S$ into equations
(\ref{eqn:masseasy}) and (\ref{eqn:voleasy}) yields
\begin{equation}
M = \frac{r_h^4}{4 l^4}, \quad %
V = \frac{8\pi r_h^4}{13 l^2}  \label{z3d3Gim}
\end{equation}
in agreement with their (quasilocal) mass  \cite{Gim2014}.
From the volume in (\ref{z3d3Gim}) we find that the isoperimetric parameter is
\begin{equation*}
	\mathcal{R} = 4 \sqrt{\frac{\pi}{13}} \frac{r_h}{l}
\end{equation*}
and therefore, the reverse isoperimetric inequality is again violated by those black holes for which $r_h$ is
sufficiently small.

\subsection{$z=D$, $k=0,1$}
\label{genlif}
We can now consider a general exact Lifshitz solution for Einstein-Dilaton-Maxwell theory
presented in \cite{Tarrio2011}  where the action is written
\begin{align}
S = \frac{1}{16 \pi G_{D}} \int d^{D}x \sqrt{-g} \bigg[ R - 2 \Lambda - \frac{1}{2} (\partial \phi)^2 - \frac{1}{4} \sum^{N}_{i=1} e^{\lambda_{i} \phi} F_{i}^2 \bigg]
\label{eqn:TarrioAction}
\end{align}
where $N \; \; U(1)$ gauge fields coupled to the scalar are considered, and the cosmological constant is fixed as
\begin{equation*}
\Lambda = - \frac{(D + z - 2)(D  + z - 3)}{2 l^2}.
\end{equation*}
The general solutions are
\begin{equation}
ds^2 = -\left(\frac{r}{l}\right)^{2z} f(r) dt^2 + \frac{l^2 dr^2}{f(r) r^2} dr^2 + r^2 d\Omega^2_{k,D-2}
\label{Fmtr}
\end{equation}
where
\begin{align}
f(r) &= k \left( \frac{ D - 3}{D + z - 4} \right)^2 \frac{l^2}{r^2}+ 1 - m r^{2 - D - z} \nonumber\\
 & \qquad + \sum_{n = 2}^{N-k} \frac{q_{n}^2 \mu^{- \sqrt{\frac{2(z-1)}{(D-2)}}} l^{2z}}{2(D-2)(D+z-4)} r^{-2(D+z-3)}
\label{Fmtrf}
\end{align}
and
\begin{align}
A'_{t,1} =& l^{-z} \sqrt{2(D+z-2)(z-1)} \mu^{\sqrt{\frac{D-2}{2(z-1)}}} r^{D+z-3}, \\
A'_{t,n} =& q_{n} \mu^{-\sqrt{\frac{2(z-1)}{(D-2)}}} r^{3-D-z}, \\
A'_{t,N} =& l^{1-z} \frac{\sqrt{2k(D-2)(D-3)(z-1)}}{\sqrt{D+z-4}} \mu^{\frac{(D-3)}{\sqrt{2(D-2)(z-1)}}} r^{D+z-5}, \\
e^{\phi} =& \mu r^{\sqrt{2(D-2)(z-1)}}
\end{align}
where $n \in [2, N-k]$  when $N \geq 2+ k$, and a prime denotes the derivative of the vector potential; 
 the $\lambda_i$ are fixed by the Einstein equations.

For $k=0$, at least one ($N=1$) $U(1)$ gauge field is required. In this case, as the gauge and the dilaton fields diverge in order to support the Lifshitz asymptotics when $r \rightarrow \infty$, the metric does not possesses a charge and yields uncharged Lifshitz black hole solutions.
For $N \geq 2$, the extra gauge field converges as $r \rightarrow \infty$ and in this case the extra $U(1)$ charges appear in the metric (\ref{Fmtr}).
This corresponds to a charged Lifshitz black hole solution.
For $k=1$, at least two ($N=2$) $U(1)$ gauge fields are necessarily required: one to support the Lifshitz asymptotics with the dilaton field
and the other to sustain the $S^{D-3}$ topology, namely the near-horizon geometry given by $AdS_{2} \times S^{D-2}$.
For $N=3$, the extra gauge charge appears in the metric (\ref{Fmtr}) and this leads to the charged Lifshitz black hole solution.
For $k=-1$, there is an imaginary charge density for $z \neq 1$ and so the hyperbolic case is only
allowed for $z=1$. Thus we confine ourselves to the $k=0$ and $k=1$ cases.

\subsubsection{Uncharged Solutions}

 The uncharged solution takes the form (\ref{Fmtr}) with all $q_n=0$, so that
\begin{equation}\label{metfUS}
f(r) = 1 - m r^{2 - D - z} + k \left( \frac{ D - 3}{D + z - 4} \right)^2 \frac{l^2}{r^2}
\end{equation}
where the one $U(1)$ (for $k=0$) and the two $U(1)$'s (for $k=1$) fields are used to fix the horizon geometry and  Lifshitz asymptotics.

The entropy for such a black hole, given by the Bekenstein-Hawking formula, is
\begin{equation}\label{SLifUS}
S = \frac{\omega_{k,D-2}}{4 G_D} r_h^{D-2}
\end{equation}
where $\omega_{k,D}$ is the $D-$dimensional area
of the unit surfaces at fixed $(t,r)$ slices,
the temperature found via the periodic Wick method is
\begin{equation}\label{TLifUS}
T = \frac{r_h^z}{4 \pi l^{1 + z}} \left[ (D + z - 2) + k \frac{(D-3)^2}{D + z - 4} \frac{l^2}{r_h^2} \right],
\end{equation}
 and the pressure becomes
\begin{equation}
P = \frac{(D+z-2)(D+z-3)}{16 \pi G_{D} l^2}.
\end{equation}

 It is straightforward to employ our thermodynamic approach to this case.  Using (\ref{eqn:smarrgen}) with equations
 (\ref{SLifUS}) and (\ref{TLifUS}) we find
\begin{align}
M = & \frac{(D-2) \omega_{k,D-2}}{16 \pi G_D}  \left( \frac{r_h^{D + z -2}}{l^{z + 1}} \right) \nonumber \\
    &+ \frac{k (D-3)^2 (D-2) \omega_{k,D-2}}{16 \pi G_D (D+z-4)^2} \left(\frac{r_h^{D+z-4}}{l^{z - 1}}\right) \label{MLifUS}\\
V = & \frac{(D-2)(z + 1) \omega_{k, D-2} }{ 2 G_D (D+z-2)(D+z-3)} \left(\frac{r_h^{z+D-2}}{l^{z-1}}\right) \\
    &+ \frac{k (D-3)^2 (D-2) (z-1) \omega_{k,D-2} }{ 2 G_D (D+z-4)^2 (D+z-2)(D+z-3)} \left( \frac{r_h^{D+z-4}}{l^{z-3}} \right)
    \nonumber
\end{align}
Notice that the $k$ terms break the scaling of $r_h$ and $l$ such that the assumption
(\ref{eqn:easytempcondition}) no longer holds.
It is straightforward to check that when $k=0$ these thermodynamic values are also consistent with the reduced Smarr formula (\ref{eqn:othersmarr})  (also used in \cite{Liu2014b} when $q=0$)
and the   thermodynamic first law (\ref{eqn:firstlaw}). 

Solving $f(r_h)=0$ for $r_h$ and inserting the result into (\ref{MLifUS}), we obtain
\begin{equation}
\label{eqn:thermomassF}
M = \frac{\omega_{k,D-2}}{16 \pi G_D} m l^{-1 -z} (D - 2)
\end{equation}
for the  thermodynamic mass. 
This result agrees with \cite{Tarrio2011} and \cite{Liu2014b}; 
the former independently calculated the mass by using a Komar integral with the
black hole solution (\ref{Fmtr}),
subtracting the value from the thermal case ($m = 0$ and $q=0$), 
while the latter reference employed the Wald formula  for planar solutions.
In the latter case we note that
only $k=0$ solutions were considered and therefore the reduced Smarr relation
(\ref{eqn:othersmarr}) was used.

\subsubsection{Charged Solutions}

From the metric (\ref{Fmtr}), we directly read off the thermodynamic variables in the charged scenario:
\begin{align}
T =& \frac{r_h^z}{4 \pi l^{z+1}} \bigg( (D+z-2) + k \frac{l^2 (D-3)^2}{r_h^2(D+z-4)} \nonumber\\
 &- \sum_{n = 2}^{N-k} \frac{q_{n}^2 \mu^{- \sqrt{\frac{2(z-1)}{(D-2)}}} l^{2z}}{2(D-2)} r_h^{-2(D+z-3)} \bigg), \label{Ftemp} \\
S =& \frac{\omega_{k,D-2}}{4 G_D} r_h^{D-2}, \label{Fentropy}\\
P =& \frac{(D+z-2)(D+z-3)}{16 \pi G_{D} l^2} \label{Fprs} .
\end{align}

 In this case, the $U(1)$ field is coupled to the scalar and so the total charge is  
\begin{equation}
Q_{i} = \frac{1}{16 \pi G_{D}} \int e^{\lambda_{i} \phi} * F = \frac{q_{i} \omega_{k,D-2} l^{z-1}}{16 \pi G_{D}}.
\end{equation} 
Inserting (\ref{Ftemp}) - (\ref{Fprs}) into the Smarr formula (\ref{eqn:smarrgenwq}) and the first  law (\ref{eqn:firstlawcharged}),
we obtain
\begin{align}
	M = & \frac{(D-2)\omega_{k, D-2}}{16 \pi G_{D}} \bigg[ \bigg(1 + k \frac{(D-3)^2 l^2}{(D+z-4)^2 r_h^2} \bigg)  l^{-z-1} r_{h}^{D+z-2} \nonumber\\
& + \sum^{N-k}_{n=2} \frac{q_{n}^2 \mu^{-\sqrt{\frac{2(z-1)}{(D-2)}}}}{2(D-2)(D+z-4)} l^{z-1} r_{h}^{4-D-z} \bigg],
\label{Fmass}\\
V = & \frac{(D-2) \omega_{k, D-2}}{(D+z-3)(D+z-2)} \bigg[ \bigg( \frac{(z+1)}{2} \nonumber\\
& + k \frac{(D-3)^2 (z-1)l^2}{2(D+z-4)^2 r_h^2} \bigg) l^{1-z}r_{h}^{D+z-2} \nonumber\\
& - \sum^{N-k}_{n=2} \frac{(z-1) q_{n}^2 \mu^{-\sqrt{\frac{2(z-1)}{(D-2)}}}}{4(D-2)(D+z-4)}l^{z+1} r_{h}^{4-D-z} \bigg],
\label{Fvol}\\
 \Phi_{n} = &   -\frac{q_n \; \mu^{-\sqrt{\frac{2(z-1)}{(D-2)}}}}{{(D+z-4)}}r_{h}^{4-D-z}.
\label{Fphi}
\end{align}
for $n=2$ through $n=N-k$.

 When $k=0$ these results agree with the reduced  Smarr
formula (\ref{eqn:simplifiedSmarr}) and the first law
(\ref{eqn:firstlawcharged}).  From the metric (\ref{Fmtr}) with
(\ref{Fmtrf}), we can read off dimensionality of $l$ in the $TS$ term.
Substituting  $\alpha = z-1$ in (\ref{eqn:simplifiedSmarr}), the Smarr
formula for $N \; U(1)$ fields takes the form
\begin{equation}\label{redsmarr3}
M = \frac{D-2}{D+z-2} T S +   \sum_{i=2}^{n} \frac{D+z-3}{D+z-2} \Phi_{i} Q_i,
\end{equation}
and the first thermodynamic law (\ref{eqn:firstlawcharged}) is satisfied.
This Smarr relation agrees with the one in \cite{Liu2014b}.

The mass in the charged case has the same value as the uncharged case, equation
(\ref{eqn:thermomassF}), and our results in equation (\ref{Fphi}) for the gauge potential  
and the charge $Q$ are consistent with the results in \cite{Tarrio2011} and   
\cite{Liu2014b} under the redefinition of charge parameter $q = \sqrt{(D-2)(D+z-4)/2} \; q_{L}$, 
where $q$ is the charge parameter in this paper and $q_{L}$ is the charge parameter in \cite{Liu2014b}.

\begin{figure}
		\centering
		\includegraphics[scale=0.6]{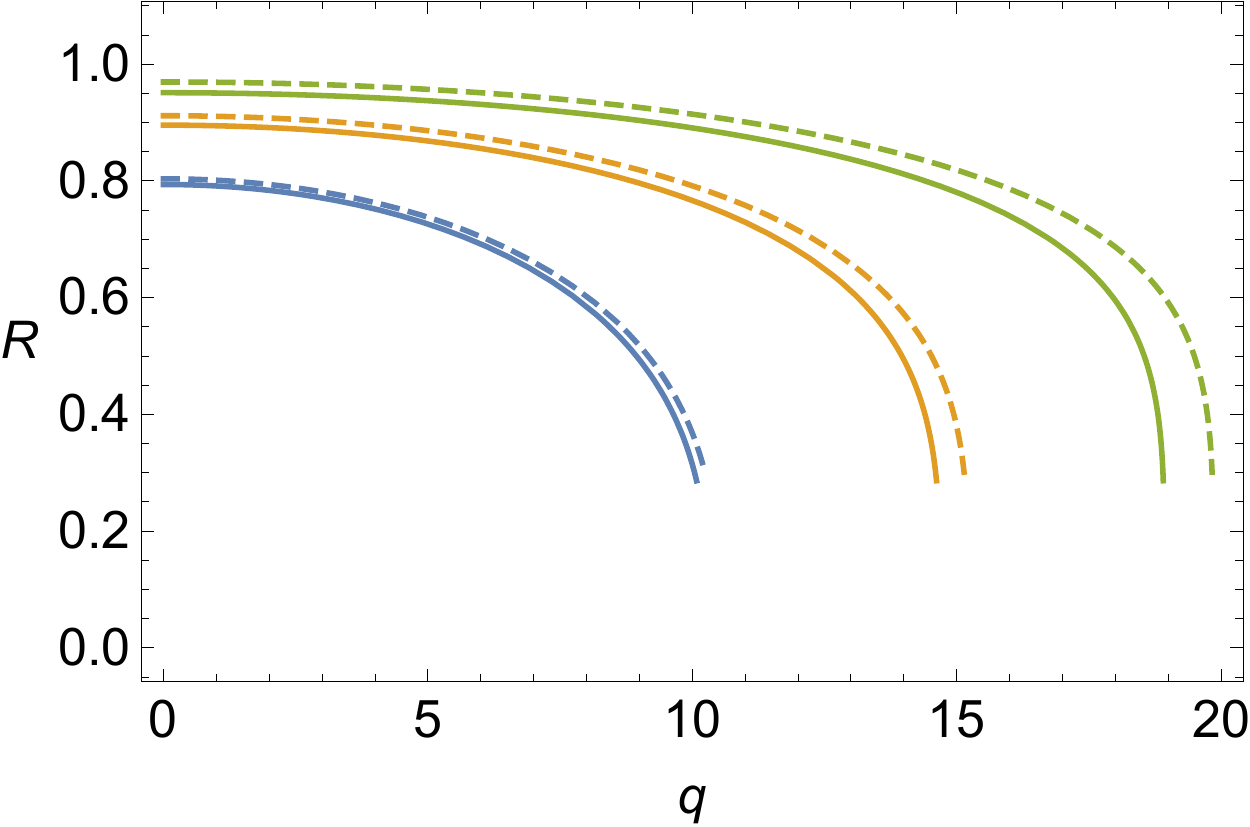}
		\caption{Plot of $\mathcal{R}$ versus the charge parameter $q$ depending on $z$ with $r_h=1$ and $l=1$ fixed. From bottom to top, $D=4,5,$ and $6$. The solid lines are for $k=0$ with $N=2$ and the dashed lines are for $k=1$ with $N=3$.}
		\label{fig:FQvsR}
\end{figure}

The thermodynamic volume (\ref{Fvol}) yields
\begin{align}
\mathcal{R}= & \bigg[ \frac{1}{4(D+z-3)(D+z-2)} \bigg\{2 (D^2-3D+2) \bigg( (z+1) \nonumber\\
& + k \frac{(D-3)^2(z-1)l^2}{(D+z-4)^2 r_h^2} \bigg) l^{1-z}r_{h}^{z-1} \nonumber\\
&- \sum^{N-k}_{n=2} \frac{(D-1)(z-1) q_n^2 \mu^{-\sqrt{\frac{2(z-1)}{(D-2)}}}}{(D+z-4)} l^{z+1} r_h^{5-2D-z} \bigg\} \bigg]^{\frac{1}{D-1}}
\end{align}
for the isoperimetric ratio, depicted in Fig. \ref{fig:FQvsR}. We see that 
 $\mathcal{R}<1$ for any value of $Q$, in strong violation of the reverse isoperimetric inequality \cite{Cvetic2010}. 

\subsection{z=2, D=4, $k=-1$}
\label{mannsol}

The  first exact Lifshitz black hole discovered was the ``topological'' black hole solution,  with a hyperbolic horizon ($k=-1$)
\cite{Mann2009} and metric function (\ref{eqn:metric}) with
\begin{equation}\label{z2TBH}
f(r) =g(r) = 1 - l^2/2 r^2
\end{equation}
in $z=2$, $D=4$.
It follows from the action
\begin{align}
S = & \int d^4 x \sqrt{-g} \bigg( R - 2 \Lambda - \frac{1}{4} F_{\mu \nu}F^{\mu \nu} - \frac{1}{12} H_{\mu \nu \tau} H^{\mu \nu \tau} \nonumber\\
& - \frac{C}{\sqrt{-g}} \epsilon^{\mu \nu \alpha \beta} B_{\mu \nu} F_{\alpha \beta} \bigg)
\end{align}
where $F= d A$, $H=dB$, and the cosmological constant and the coupling constant $C$ are fixed as
\begin{equation}
\Lambda = - \frac{z^2 + z +4}{2 l^2}, \; \; \; \; 2 z = (Cl)^2 .
\label{Glmbd}
\end{equation}
The gauge fields are
\begin{equation}
	F_{tr} = 2 \frac{r}{l^2}, \; \; H_{r \theta, \phi} = 2 r \sinh \theta .
\label{eqn:Gmtr}
\end{equation}

The cosmological constant in this case is $\Lambda = - 5/l^2$. From the metric (\ref{eqn:Gmtr}) and the cosmological constant (\ref{Glmbd}), the thermodynamic variables are calculated as
\begin{align}
T = \frac{1}{4 \pi l}, \; \; \; S = \frac{r_h^2}{4} \omega_{-1,2} = \frac{l^2}{8} \omega_{-1,2}, \; \; \; P = \frac{5}{8 \pi l^2}
\end{align}
where $r_h=\frac{l}{\sqrt{2}}$ is  found from $f(r_h)=0$.

Various attempts have been made
to find a mass for this black hole. Notably,
the Brown-York and Hollands-Ishibashi-Marolf masses
were computed in \cite{Mann2011a}.
Converting those values to the conventions used here, they
find that
\begin{equation*}
	{\cal{M}}_{BY} = \frac{\hat{l} \tilde{\text{Vol}}_{\Sigma}}{8 \kappa^2} = \frac{l   \omega_{-1,2}}{32 \pi}
\end{equation*}
while for minimal coupling
\begin{equation*}
	{\cal{M}}_{HIM} = \frac{3 \hat{l} \tilde{\text{Vol}}_{\Sigma}}{32 \kappa^2} = \frac{3 l   \omega_{-1,2}}{128 \pi}
\end{equation*}
and using an extended action (where additional surface terms were added to allow independent variations
of the asymptotic behaviour of the metric function and the gauge field),
\begin{equation*}
	{\cal{M}}^{(E)}_{HIM} = \frac{l \omega_{-1,2}}{128 \pi} .
\end{equation*}

We could also use the $m=0,z=2,D=4$ case of the previously studied
dilaton solution in section \ref{genlif} to guess at the mass for
the solution (\ref{z2TBH}).
The dilaton solution would have $\mathcal{M}_K = 0$ in this case.
 Of course, this is at best a guess since this solution requires
the dilaton to exist; however, the methodology is similar.
 
 We can classify all of our thermodynamically inspired approaches towards a mass as follows.
For this solution the horizon radius  is dependent only on the cosmological length scale. There are no other independent parameters, and so in general the thermodynamic volume will not be uniquely determined without further assumptions.  
However, simple geometric and dimensional considerations imply that  the expression for the mass must have the form
\begin{equation}
M = \frac{\hat{m} l \omega_{-1,2}}{32 \pi}
\end{equation}
where $\hat{m}$ is a dimensionless constant whose particular value depends on the assumptions employed.  The Smarr relation then implies
\begin{equation*}
	 \frac{\hat{m} l \omega_{-1,2}}{32 \pi} = (D-2) TS - 2 PV = 2 \left( \frac{l \omega_{-1,2}}{32 \pi} \right) - 2 \frac{5}{8 \pi l^2} V
\end{equation*}
yielding
\begin{equation}
	V = (2-\hat{m}) \frac{l^3}{40} \omega_{-1,2} \label{vol-0}
\end{equation}
for the thermodynamic volume.
To proceed further we shall consider several approaches as discussed in section \ref{dependentlength}.

One is to fix the mass by some criterion.  Several possibilities suggest themselves:
the Brown-York mass ($\hat{m}=1$), the  Hollands-Ishibashi-Marolf (HIM) mass ($\hat{m}=3/4$), its value from
the extended action ($\hat{m}=1/4$),  and zero mass.  The volume can then be determined from each using (\ref{vol-0}).

Another is to  introduce  a fictitious mass parameter.
This entails a natural extension of the metric  function to 
\begin{equation}
f(r) = 1 - \frac{l^2}{2 r^2} + \tilde{m} \frac{l^p}{r^p}
\label{eqn:metanz1}
\end{equation}
where $p$ is chosen so as to yield a desired falloff rate.  The thermodynamic approach is employed and the limit
$\tilde{m}\to 0$ is taken at the end of the calculation. Two values of $p$ naturally suggest themselves: (a) $p=D-1=3$, thereby requiring the same falloff 
as the Schwarzschild case -- this yields  $\hat{m}=1/4$ --
and (b) $p=D+z-2=4$, agreeing with the form of the mass term in the dilatonic
general solution \cite{Tarrio2011}, and yielding  $\hat{m}=0$.

A third approach is to simply use the expression $S = \frac{r_h^2}{4} \omega_{-1,2}$ for the entropy and proceed as though
$r_h$ and $l$ were independent quantities, setting $r_h\to l/\sqrt{2}$ at the end of the calculation.  This gives $\hat{m}=1$.
A final possibility is to fix $V = 0$, yielding $\hat{m} = 2$.

Note that basic physical considerations imply that $0\leq \hat{m} \leq 2$ in
order for both mass and thermodynamic volume to remain positive.

  \begin{center}
  \begin{table}
  \caption{Values of the mass parameter $\hat{m}$ for various \\
  methods of obtaining mass for the $z=2$ solution (\ref{z2TBH})}
  \label{tab:himby}
        \begin{tabular}{l | c}
                  method	& $\hat{m}$ \\
            \hline
            Brown-York & 1 \\
            HIM & 3/4 \\
            HIM-extended & 1/4 \\
            $p =  (D-3)$ & 1/4 \\
            $p =  (D+z-2)$ & 0 \\
            $M = 0$ & 0 \\
            $V = 0$ & 2 \\
            $r_h,l$ independent & 1
                    \end{tabular}
    \end{table}
  \end{center}

In the dilaton theory of
equation (\ref{eqn:TarrioAction}), we would
have seen that $M=0$ and $V = l^3 \omega_{-1,2}/48$ 
upon substitution of $D=4$ and $z=2$ into the modification of equation (\ref{MLifUS}), 
after redefining $l$ such that the metric function (\ref{Fmtrf}) agrees with equation (\ref{z2TBH}).
Though the two cases are not identical (the differing cosmological constant means the volume
is not $V = l^3 \omega_{-1,2}/20$ in \ref{vol-0} as when $\hat{m} = 0$), it is
plausible that mass should behave in a similar way in both solutions since the only
thermodynamic difference is due to a different cosmological constant.

The Brown-York and Hollands-Ishibashi-Marolf masses provide justification for two additional
approaches. We find that the Brown-York mass, evaluated on an asymptotic surface, agrees with
the approach of assuming independence of $r_h$ and $l$ in the derivations of the entropy and
temperature. The HIM mass with an extended action (to allow for independent variation of the Proca
field) agrees with the AdS-Schwarzschild-inspired $(D-1)$ falloff of a fictitious mass term.

These results are tabulated in Table \ref{tab:himby}.

\subsection{z=4, D=4,  {$k=1,0$ and $-1$}}
\label{chargedpeet}

Here we examine
the Lifshitz black hole discovered in  \cite{Bertoldi2009} and expanded to
accommodate a Maxwell field in \cite{Brynjolfsson2010b} when $z=4$ and $D=4$. The action with a massive vector field and a Maxwell field is
\begin{align}
S = \frac{1}{16 \pi G} \int d^4 x \sqrt{-g} (R - 2 \Lambda - \frac{1}{4}H^2 - \frac{m^2}{2} B^2 - \frac{1}{4}F^2)
\end{align}
where $\Lambda = -12/l^2$ and the metric and the vector field solutions are  
\begin{align}
\label{eqn:maxwellVectorPotentialPeet}
& B_{t} = \sqrt{\frac{3}{2}} \frac{r^4}{l^4} f(r), \; \; \; \; A_{t} = {\frac{r^2}{l^3} q},\\
& ds^2 = - \frac{r^{2z}}{l^{2z}} f(r) dt^2 + \frac{l^2 dr^2}{r^2 f(r)} + r^2 d \Omega_{k}^2
\end{align}
where
\begin{equation}
f(r) = 1 + a \frac{k {l^2}}{r^2} - b \frac{k^2 {l^4}}{r^4} - \frac{q^2 l^2}{2 r^4}
\label{eqn:Hmtr}
\end{equation}
with $a = \frac{1}{10}$ and  $b=\frac{3}{400} = \frac{3 a^2}{4}$. When $ {q}=0$,
the metric in equation (\ref{eqn:Hmtr}) is a pure Lifshitz spacetime for $k=0$, 
a black hole with a spherical horizon for $k=1$, and a topological black hole with a hyperbolic horizon for $k=-1$.
When $Q \neq 0$, charged black hole solutions are present for all $k$.

The temperature, entropy, and pressure are straightforwardly calculated to be
\begin{align}
& T = \frac{1}{2 \pi l} \bigg( -a k \frac{r_{h}^{2}}{l^{2}} {+ 2 b k^{2}} + \frac{q^{2}}{l^{2}} \bigg), \label{Htemp}\\
& S = \frac{r_h^2}{4} \omega_{k,2}, \; \; \; \; \; P = \frac{3}{2 \pi l^2}.
\label{Htspwq}
\end{align}
and the Maxwell charge is
\begin{equation}
Q = \frac{1}{4 \pi} \int *F = \frac{q}{2 \pi} \omega_{k,2}.
\label{eqn:maxwellPeet}
\end{equation}
where $r_h$ is obtained from $f(r_h)=0$ in (\ref{eqn:Hmtr})
\begin{equation}
r_h = { \sqrt{l \left( -\frac{a k l}{2} +\sqrt{\frac{1}{4} a^2 k^2 l^2+b k^2 l^2+\frac{q^2}{2}}\right)} }.
\label{Hhr}
\end{equation}

This is the first solution we encounter where there is no mass parameter as well as no
agreed-upon derivation of mass. As in section (\ref{mannsol}),
the length and horizon radius of this black hole (when uncharged) are dependent. Therefore,
the Smarr relation and the first law become degenerate in this case, and so mass and
volume are not unique.

For the same reasons as the previous case, when $Q=0$,  geometric and dimensional considerations imply
\begin{equation}
M = \frac{\hat{\mathfrak{m}} l \omega_{k,2}}{800 \pi}
\end{equation}
where $\hat{\mathfrak{m}}$ is a dimensionless constant.  The Smarr relation then implies
\begin{equation*}
	 \frac{\hat{\mathfrak{m}} l \omega_{k,2}}{800 \pi} = (D-2) TS - 2 PV =
	 \frac{\omega_{k,2} r_h^2}{4\pi l^3}\left(-k a r_h^2 +2b k^2 l^2\right)   -  \frac{3}{\pi l^2} V
\end{equation*}
yielding
\begin{equation}
	V =
	({5|k|- 4 k}- 10 \hat{\mathfrak{m}}) \frac{l^3}{24000} \omega_{k,2} \label{vol-Q0}
\end{equation}
for the thermodynamic volume, where $k=\pm 1$.

When $Q\neq 0$ a much broader range of possibilities emerges for the form of $M$ based only on geometrical considerations, since there are now two length scales present.  The Smarr relation (\ref{eqn:smarrgenwq}) and charge conjugation invariance, however, suggest
\begin{equation}\label{Manz1}
M = \frac{\hat{m} l \omega_{k,2}}{800 \pi} + \frac{\hat{w} r^2_h q^2  \omega_{k,2}}{4 l^3 \pi}
\end{equation}
from the form of $\Phi Q$  and the independence of $Q$ (\ref{temp3}). 

In general $\hat{m}$ and $\hat{w}$ could be dimensionless functions of both $q/l$ and $r_h/l$.   
We shall remain open to this possibility in what follows.
Taking (\ref{Manz1}) as an ansatz implies
\begin{align}
\label{eqn:SmarrQTerms}
& \frac{\hat{m} l \omega_{k,2}}{800 \pi} + \frac{\hat{w} r^2_h q^2  \omega_{k,2}}{4 l^3 \pi} \\
& \quad=
\frac{\omega_{k,2} r_h^2}{4\pi l}  \bigg( 2b k^2 + \frac{q^2}{l^2}- k \frac{a r_h^2}{ l^2} \bigg)
-  \frac{3}{\pi l^2} V  + \frac{\hat{w} r^2_h q^2  \omega_{k,2}}{2 \pi l^3 }\nonumber
\end{align} 
from the Smarr relation, with
\begin{equation}
\Phi  = \frac{\partial M}{\partial Q}= \hat{w} \frac{r_{h}^{2}}{l^{3}} q 
\end{equation}
and so we obtain
\begin{equation}
V = \left(3 k^{2} \frac{ r_{h}^2}{l^{2}} - 20 k \frac{r_{h}^4}{l^{4}}- \hat{m} \right) \frac{ l^{3 }\omega _{k,2}}{2400} +( \hat{w}+1)  \frac{q^{2} r_{h}^2  \omega _{k,2}}{12 l}
\end{equation}
for the general form for the thermodynamic volume.

We first approach the problem of determining $\hat{m}$ and $\hat{w}$ using a fictitious mass term.
As an example, we can work out the case with scaling
$\tilde{m} (l/r)^{(D+z-2)}$.
In order to find a unique thermodynamic volume and mass, we can again manually separate horizon radius
and lengthscale  as in section (\ref{dependentlength}).
The most obvious choice would be to add a parameter that appears in a manner similar to the mass of
known Lifshitz black holes, namely a term that scales like $(l/r)^{(D+z-2)}$ in $f(r)$.
Our ansatz is then
\begin{equation*}
		f(r) = 1 + k \frac{l^2}{10 r^2} - k^2 \frac{3 l^4}{400 r^4} - \frac{q^2 l^2}{2 r^4}  + m \frac{l^6}{r^6}.
\end{equation*}
and our algorithm returns for the $q=0$ case
\begin{align}
M &= \omega_{k, 2} \left( - \frac{3 k^2 r_h^2}{3200 \pi l} + \frac{k r_h^4}{80 \pi l^3} + \frac{r_h^6}{8 \pi l^5}\right) \\ %
V &= \omega_{k, 2} \left( -\frac{ k^2 l r_h^2}{3200} + \frac{k r_h^4}{80 l} + \frac{5 r_h^6}{{24} l^3} \right)
\end{align}

Taking $m=0$ and substituting $r_{h}$ in (\ref{Hhr}) yields $\hat{m} = 0$. Then for $k=1$ the thermodynamic mass and volume become 
\begin{equation*}
M = 0 \qquad %
V = \frac{ \omega_{1,2} l^3}{24000} = \frac{ \pi l^3 } {6000}
\end{equation*}
with $r_h = l/2 \sqrt{5}$ and  the reverse isoperimetric inequality is
\begin{equation*}
\mathcal{R} = \frac{1}{\sqrt{5}} \cdot \left( \frac{1}{8} \right)^{1/3} = \frac{1}{2 \sqrt{5}} \approx 0.2236.
\end{equation*}
For $k=-1$ the thermodynamic mass and volume become
\begin{equation*}
M = 0 \qquad %
V = \frac{ 3 \omega_{-1,2} l^3}{8000}
\end{equation*}
 with $r_{h} = \frac{1}{2} \sqrt{\frac{3}{5}} l$ and  the reverse isoperimetric inequality is
\begin{equation*}
\mathcal{R} =\frac{\sqrt[6]{3}}{2 \sqrt{5}} \approx 0.2685.
\end{equation*}
 
Adding charge as above, we find using the fictitious mass approach that
\begin{align}
\label{FtmassH}
M &= \omega_{k, 2} \left( - \frac{3 k^2 r_h^2}{3200 \pi l} + \frac{k r_h^4}{80 \pi l^3} %
+ \frac{r_h^6}{8 \pi l^5}\right) %
- \frac{\omega_{k,2} q^2 r_h^2}{16 \pi l^3} \\  %
V &= \omega_{k, 2} \left( -\frac{ k^2 l r_h^2}{3200} + \frac{k r_h^4}{80 l} %
+ \frac{5 r_h^6}{{24} l^3} \right) %
- \frac{\omega_{k,2} q^2 r_h^2}{16 l}.
\end{align}

Inserting the value (\ref{Hhr}) for the horizon radius, we obtain  
\begin{align}
\hat{m}=50 \frac{q^{2} r_{h}^{2}}{l^{4}}, \qquad \hat{w} = -\frac{1}{4}, 
\end{align}
yielding $M=0$ for all $k$. The thermodynamic volume and electric potential become 
\begin{align}
&V = \frac{ (kl-2 x) \left(-2 k^2 l^2+ k l x -50 q^2\right)}{24000}\omega _{k, 2},\\
&\Phi = -\frac{q r_h^2}{4 l^3},
\end{align}
where $x = \sqrt{k^2 l^{2} + 50 q^{2}}$; 
these satisfy the Smarr relation
\begin{align}
0 = 2 T S - 2 P V + \Phi Q 
\end{align}
as well as the first law of thermodynamics
\begin{align}
&{0} = T dS + V dP + \Phi  dQ,
\end{align}
which simplifies to  
\begin{align*}
&0 = T S + \Phi Q \\
& {0} =  T dS + \Phi  dQ
\end{align*}
upon setting $k=0$.

\begin{figure}
		\centering
		\includegraphics[scale=0.8]{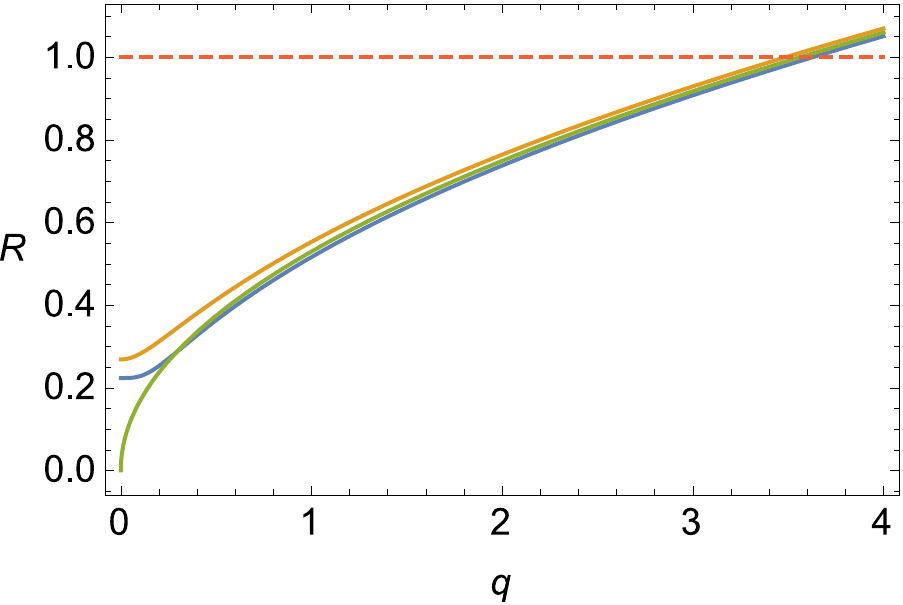}
		\caption{Plot of $\mathcal{R}$ versus charge parameter $q$ for $k=1, -1$ and $0$, which correspond to blue, orange, and green.}
		\label{fig:chargedPeetIso}
\end{figure}

Computing $\mathcal{R}$, we obtain
 \begin{equation}
 \mathcal{R} =\frac{1}{2\sqrt{5}l} \sqrt[3]{-\frac{l \sqrt{l (2 y-k l)} \left(2 k^2 l^2-k l y+50 q^2\right)}{k l-2 y}}
\end{equation}
where $y=\sqrt{k^{2} l^{2} + 50 q^{2}}$. This is depicted in terms of $q$ in Figure \ref{fig:chargedPeetIso}.  It is clear
that the reverse  isoperimetric inequality is initially violated but as $q$ increases,
it is eventually satisfied. The asymptotic behaviour is $\mathcal{R} \sim \sqrt{q}$.

These solutions correspond to the $p = (D+z-2)$ row in Table \ref{tab:peet1}.  We pause to remark
that had we employed the ansatz
\begin{equation}\label{Manz2}
M = \frac{\hat{\upsilon} r_h^6 \omega_{k,2}}{48 \pi l^5} + \frac{\hat{\zeta} k r_h^4 \omega_{k,2}}{160 \pi l^3} + \frac{3 \hat{\phi} k^2 r_h^2 \omega_{k,2}}{1600 \pi l} + \frac{\hat{\phi} r^2_h q^2  \omega_{k,2}}{8 l^3 \pi}
\end{equation}
we would have obtained the same results as above, but with $\hat{\upsilon} = 6, \hat{\zeta} = 2, \hat{\phi} = -1/2$
all being constants.

We can also approach this solution by making a comparison with the RN-AdS black holes in section III-A. Recalling the metric functions for  the RN-AdS (\ref{eqn:fr}) and  Lifshitz  (\ref{eqn:Hmtr}) solutions
\begin{align*}
f_{\text{RN}}(r) &= 1 + k \frac{l^{2}}{r^{2}} - \frac{2 m l^{2}}{r^{3}} + \frac{q^{2} l^{2}}{r^{4}},\\
f_{\text{Lif}}(r) &= 1 + k \frac{a l^{2}}{r^{2}} -  \frac{b k^{2} l^{4}}{r^{4}} - \frac{q^{2}l^{2}}{2r^{4}} .
\end{align*} 
For both metric functions above,   the first term is due to a cosmological constant (which in the Lifshitz case depends on the Proca charge), the second term determines the horizon geometry, and the fourth term is generated by a Maxwell charge.  In applying our approach to the RN-AdS case, we assume that
$r_{h}, l$, and $q$ are independent, so that $f_{\text{RN}}(r_{h})=0$ implies $m=m(r_{h},l,q)$. We shall take the same approach for the Lifshitz case with the parameter $b$, 
 assuming $r_{h}, l$, and $q$ are independent, and setting $b=b(r_{h},l,q,a)$ from $f_{\text{Lif}}(r_{h})=0$.

It is also interesting to note that the $q^{2}$ term appears with opposite sign in the Lifshitz case; furthermore the electric potential $A_t$ differs in its  $r$-dependence from the RN-AdS solution.
The sign difference can be understood from the electric potential:
when $l$ is fixed, the RN-AdS electric potential falls off with $1/r$ and approaches zero asymptotically. 
On the other hand, in the Lifshitz solution the electric potential grows with $r^{2}$, 
so the electric charges preferably tend to move towards the origin until their electric potential is balanced with the gravitational force. 
Thus the electric charges contribute positively towards the gravitational energy of the system. 

 For $k=1$ and $-1$, we can eliminate $b$ in terms of the horizon length,
 setting $b=\frac{a r_{h}^2}{k l^2}+\frac{r_{h}^4}{k^2 l^4}-\frac{q^2}{2 k^2
 l^2}$. By integrating $T S'(r_{h})$ with respect to $r_{h}$ we then obtain the
 thermodynamic mass. Taking the variation of this quantity and the pressure with
 respect to $l$ in (\ref{thermoV}) we then compute the thermodynamic mass and volume as 
\begin{align}
&M = \frac{(k l-2 x)^{2}(k l+4 x)}{192000 \pi l^{2}}  \omega_{k,2}, \\
&V = - \frac{(k l-20x)(k l-2x)^{2}}{576000} \omega_{k,2}, \nonumber
\end{align}
where $x=\sqrt{k^2 l^{2} + 50 q^{2}}$, with the uncharged case easily obtained by setting $q=0$.
 Interestingly the electric potential at the horizon becomes zero ($\Phi = 0$), 
$ \partial_{Q} M = \Phi = 0$.

\begin{figure}
		\centering
		\includegraphics[scale=0.8]{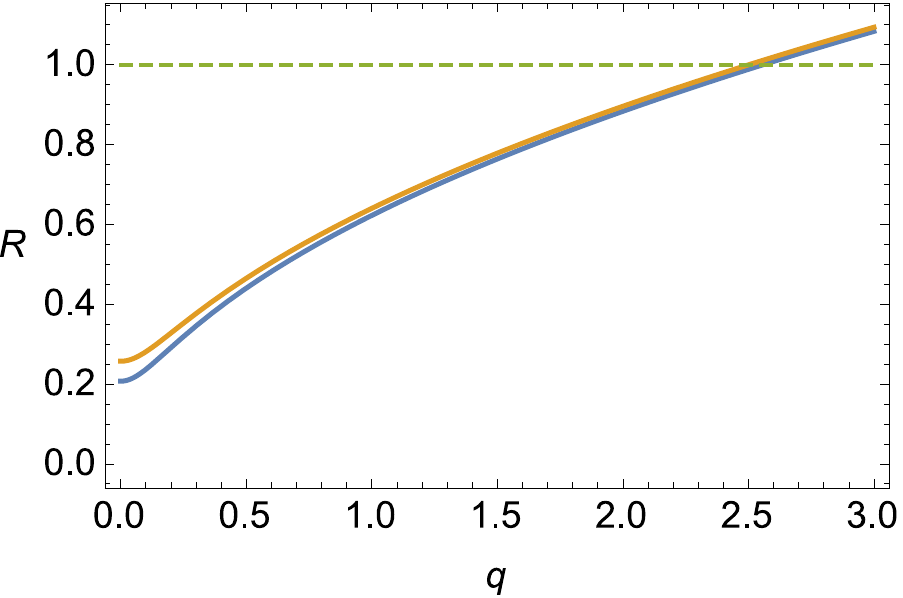}
		\caption{Plot of $\mathcal{R}$ versus the charge parameter $q$. The blue line is for $k=1$ and the orange line is for $k=-1$.}
		\label{fig:HQvsR}
\end{figure}

The isoperimetric ratios are 
\begin{equation}
\mathcal{R}=\frac{\sqrt[3]{l \sqrt{l (2 x -k l)} (20 x -k l)}}{4 \sqrt[3]{3} \sqrt{5} l}
\end{equation}
and depicted in Figure (\ref{fig:HQvsR}).

 For the $k=0$  case we integrate $T S'(r_{h})$ with respect to $r_{h}$ by setting $a=b=0$ in (\ref{Htemp}) in order to get a mass, assuming the independence of $r_h$, $l$ and $q$.  We find 
\begin{align}
M = \frac{1}{2}\Phi_{H} Q,  \; \; \; V = \frac{q^2 r_h^2}{8 l} \omega_{0,2}, \; \; \; \Phi = \frac{q r_{h}^{2}}{2 l^{3}} 
\label{Hmasswqk0}
\end{align}
using either (\ref{Manz1}) or (\ref{Manz2}), 
with $r_h = \sqrt{\frac{q l}{\sqrt{2}}}$. Here the thermodynamic mass is generated by the electric charge and its potential.

 The isoperimetric ratio is
\begin{equation}
\mathcal{R} = \frac{3^{1/3}}{2^{11/12}} \sqrt{\frac{q}{l}} = \frac{3^{5/6}}{2^{5/3}} \sim 0.787.
\end{equation}

It is straightforward to check that all thermodynamic quantities computed in this section satisfy the Smarr formula (\ref{Smarrchaged}), where for $k=0$
\begin{equation}
M = \frac{1}{3} T S +\frac{1}{3} \Phi_{H} Q.
\end{equation}
We tabulate the results of this section in
 Tables \ref{tab:peet1} and  \ref{tab:peet2}.  As a final comment, using the
 fictitious mass approach, the dimensionless constants 
in Table \ref{tab:peet2} are related to the fictitious mass term's scaling: 
$\hat{\upsilon} = p$, $\hat{\zeta} = -1 + p/2$, and $\hat{\phi} = 1 - p/4$.

\begin{center}
 \begin{table}[htp]
 \caption{Parameter values  for the ansatz (\ref{Manz1}) \\
for the $z=4$ solution (\ref{eqn:Hmtr})}
 \label{tab:peet1}
	 \begin{tabular}{l | c | c | c | c}
		method	& \; Q \; & k & $\hat{m}$ & $\hat{w}$ \\
            \hline \hline
            \multirow{2}{*}{$p=(D+z-2)$} & 0 & 1, -1 & 0 & 0 \\
            & $q$ & 1, -1, 0& $50 \frac{q^{2} r_{h}^{2}}{l^{4}}$ &  $-\frac{1}{4}$ \\ \hline
            \multirow{3}{*}{ $b=b(r_{h},l,q,a)$}   &  0 & 1 & $\frac{1}{48}$ & 0   \\
             &  0 & -1 & $\frac{9}{80}$ &  0   \\
            & $q$ & 1, -1 & $5\left(k \frac{r_{h}^{4}}{l^{4}} + \frac{40}{3} \frac{r_{h}^{6}}{l^{6}} \right)$ & 0  \\
           {$r_h,l,q$ independent} & $q$ & 0 & 0 & $\frac{1}{2}$
        \end{tabular}
    \end{table}
  \begin{table}[htp]
 \caption{Parameter values  for the ansatz (\ref{Manz2}) \\
for the $z=4$ solution (\ref{eqn:Hmtr})}
 \label{tab:peet2}
	 \begin{tabular}{l | c | c | c | c}
		method & k & $\hat{m}$ & $\hat{n}$ & $\hat{\phi}$ \\
            \hline \hline
            \multirow{1}{*}{$p=(D+z-2)$} & 1, -1, 0 & 6 & 2 & -1/2\\
	     $p =  (-D) \text{ or } (-z)$ & 1,-1,0 & -4 & -3 & 2 \\
            \multirow{1}{*}{ $b=b(r_{h},l,q,a)$}   & 1, -1 & 4 & 1 & 0   \\
           {$r_h,l,q$ independent} & 0 & 0 & -1  & 1
        \end{tabular}
    \end{table}
  \end{center}

\subsection{z = 2(D-2), $k=0$}

The final set of exact solutions we shall consider are those based off of a
$k=0$ black brane solution in \cite{Brynjolfsson2010b} and generalized to arbitrary
dimension in Pang \cite{Pang2009}.
The relevant action consists of a Proca field as well as a Maxwell field, given by
\begin{align}
S = \frac{1}{16 \pi G_{D}} \int d^D x \sqrt{-g} & \bigg( R - 2 \Lambda  - \frac{1}{4}H^2 \bigg. \nonumber \\
& \bigg. \quad - \frac{1}{2}m^2 B^2 - \frac{1}{4} F^2  \bigg)
\end{align}
where  $H$ is the Proca field strength, $H = dB$, 
 $F$ is the Maxwell field strength, and the cosmological constant is 
\begin{align}
\Lambda = - \frac{(z-1)^2 + (D-1)(z-2) + (D-1)^2}{2 l^2}.
\end{align}
It is known that the solutions take the form
\begin{align}
&B_t = \sqrt{\frac{2(z-1)}{z}} \frac{r^z}{l^z} f(r) , \; \; \; \; \; F_{rt} = q l^{1-z} r^{-D+z+1},\label{Ifldstngth} \\
&ds^2 = -\frac{r^{2z}}{l^{2z}} f(r) dt^2 + \frac{l^2}{r^2} \frac{dr^2}{f(r)} + r^2 \sum^{D-2}_{i=1} dx_{i}^2 \label{Imetric}
\end{align}
where
\begin{align}
f(r) = 1 - \frac{q^2 l^2}{2(D-2)^2 r^z},
\end{align}
and are allowed only when $z=2(D-2)$. In this solution, when $r \rightarrow \infty$ the Maxwell field strength diverges, but the part of the Proca field  associated with the Maxwell charge converges.

 We can immediately read off temperature, entropy, and pressure as
\begin{align}
\label{Itsp}
	T &= \frac{l^{5-2D} q^2}{4\pi(D-2)}, \;  \; \; \; \; \; \; \; \; S = \frac{r_h^{D-2}}{4} \omega_{0,D-2},\\
    P &= \frac{7D^2 -30D+32}{16 \pi l^2},
\end{align}
and the Maxwell charge is
\begin{equation}
Q = \frac{1}{4 \pi} \int *F = \frac{q}{4 \pi} \omega_{0,D-2}
\label{eqn:maxwellZD}
\end{equation}

The approach of classifying the scaling of fictitious mass with a function
\begin{equation}
f(r) = 1 + m \left( \frac{l}{r} \right)^p - \frac{q^2 l^2}{2 (D-2)^2 r^z}
\end{equation}
can be applied here as well, yielding 
\begin{align*}
M &= \frac{ \hat{m} \omega_{0,D-2} r_h^{3D-6}}{48 \pi l^{2D -3}} + \frac{q^2 \hat{w} \cdot \omega_{0,D-2} r_h^{D-2}}{32 (D-2)^2 \pi l^{2D-5}} \\
P V &= \frac{(2D-3) \hat{m} \omega_{0,D-2} r_h^{3D-6}}{96 \pi l^{2D -3}} + \frac{ (2D-5) q^2 \hat{w} \cdot \omega_{0,D-2} r_h^{D-2}}{64 (D-2)^2 \pi l^{2D-5}} \\
\Phi Q &= \frac{q^2 \hat{w} \cdot \omega_{0,D-2} r_h^{D-2}}{16 (D-2)^2 \pi l^{2D-5}}
\end{align*}
where $\hat{m} = p$ and $\hat{w} = 2(D - 2) - p$. 
Notably, when $p = (D+z-2)=3(D-2)$ we again find that $M = 0$. This is easier to see by combining
the above terms under the solution $q^2 = 2(D-2)^2 r_h^{2(D-2)}/l^2$:
\begin{align*}
M &= \frac{ (6(D-2) - 2p) \omega_{0,D-2} r_h^{3D-6}}{48 \pi l^{2D -3}} \\
P V &= \frac{(6 D^2 -27 D + 30 + 2p(3-D)) \omega_{0,D-2} r_h^{3D-6}}{48 \pi l^{2D -3}} \\
\end{align*}

Alternatively, we can assume that $r_h$, $l$ are independent. 
By using the relations (\ref{firstlaweqs}) and (\ref{pMovpQ}) with the mass ansatz (\ref{thermoM}), we obtain the thermodynamic mass, volume, and electric potential
\begin{align}
&M =  \frac{l^{5-2D} r_h^{D-2} q^2 \omega_{0,D-2}}{16 \pi (D-2)}=\frac{1}{2} \Phi Q , \label{Imass}\\
&V = \frac{(2D-5)l^{7-2D}r_h^{D-2}q^2 \omega_{0,D-2}}{2(D-2)^2 (7D-16)} , \label{Ivol}\\
&\Phi = \frac{l^{5-2D} r_h^{D-2} q}{2 (D-2)}. \label{Iphi}
\end{align}
With this restriction, these thermodynamic variables are consistent with the
first thermodynamic law (\ref{eqn:firstlawcharged}) and the Smarr equation (\ref{eqn:smarrgenwq}). 
Here the thermodynamic mass (\ref{Imass}) is expressed by the charge $Q$ of the system, and so the energy of spacetime represented by (\ref{Imetric}) is generated by the charge.

Since this spacetime has a planar horizon ($k=0$) we find that (\ref{Imass}) and (\ref{Iphi})  are consistent with (\ref{eqn:smarrgenwq}) and (\ref{eqn:firstlawcharged}); they also satisfy (\ref{eqn:fstlawwthpv}) and (\ref{eqn:othersmarr}), from which the reduced Smarr formula is
\begin{align}
M = \frac{1}{3} TS + \frac{1}{3} \Phi Q
\label{eqn:pangSimplSmarr}
\end{align}
with $\delta = 5-2D$ in the form of equation (\ref{eqn:simplifiedSmarr}).

\begin{figure}
		\centering
		\includegraphics[scale=0.8]{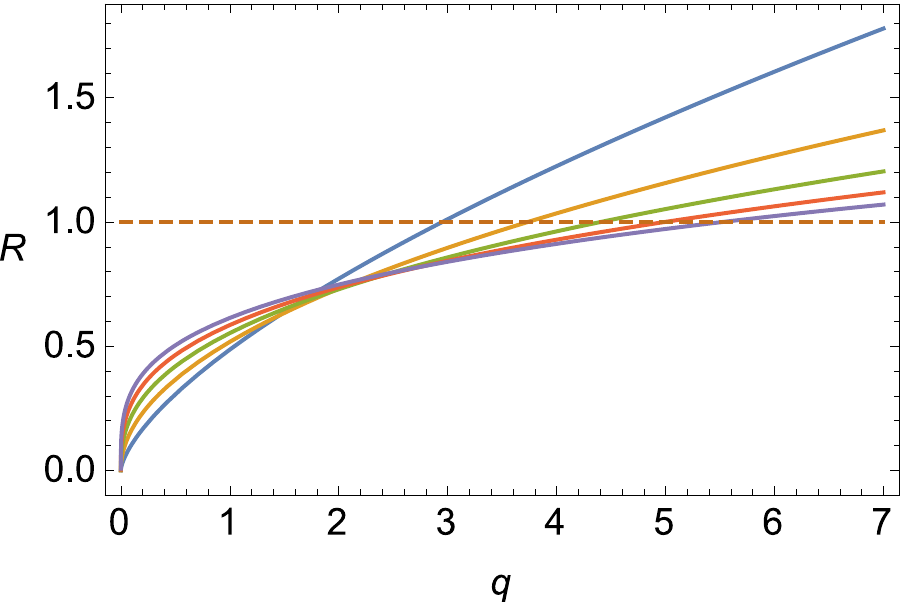}
		\caption{Plot of $\mathcal{R}$ versus the charge parameter $Q$ depending on the dimension of the spacetime $D$ with $l=1$ fixed. From top (blue) to bottom, $D=4,5,6,7$ and $8$.}
		\label{fig:IQvsR}
\end{figure}

 The isoperimetric parameter is
\begin{equation}
\mathcal{R} = \bigg( \frac{l^{7-2D}(D-1)(2D-5) q^2}{2(D-2)^2(7D-16)r_h} \bigg)^{\frac{1}{D-1}},
\end{equation}
and is plotted in Fig.~\ref{fig:IQvsR}. For each value of $D$, there is a threshold value of $q$ for which
the reverse isoperimetric inequality is satisfied ($\mathcal{R}>1$), but this threshold value increases as
$D$ increases.

Recently an attempt was made to independently compute a mass for this solution  in $D=4$ \cite{Liu2014b}.
It uses the Wald formula, implying a first law, to obtain $\mathcal{M}_{Wald} = 0$.
The Smarr relation that they use is
$0 = TS + \Phi Q$,
arising from the $k=0$ simplified version of
equation (\ref{eqn:simplifiedSmarr}), or equivalently the $M=0$ case
of equation (\ref{eqn:pangSimplSmarr}).
 
However, we find that the metric function and Proca field given in \cite{Liu2014b}
do not satisfy the  field equations. Consequently the Smarr relation requires a value of $\Phi Q$
different from equations (\ref{Iphi}) and (\ref{eqn:maxwellZD}) in order to
hold with zero mass.
We can choose $p$ such that our result either agrees with the Wald approach above
or the Maxwell $\Phi Q$ term, but not both unless we choose $M=0$ and allow $q$ to
be dependent on $l$.
As before, a $p = 0$ scaling produces $\Phi Q$ agreement (and is equivalent
to the method of assuming independence) while yielding a finite mass.
 
 Because there is no conclusive independent result for mass in this example,
the thermodynamic method cannot fully prescribe the form for either the volume or the
mass, and we leave this determination for future work.
 
\section{Conclusion}

We have shown that the mass of a Lifshitz black hole 
can indeed be understood as enthalpy, with the general Smarr formula (\ref{eqn:smarrgen})
valid for all such spacetimes.  Using this formula and the first law of thermodynamics, we are
able to determine a thermodynamic mass/enthalpy and thermodynamic volume. 
For $k=0$ these thermodynamic quantities are also consistent with the reduced forms
(\ref{eqn:masseasy}) and (\ref{eqn:simplifiedSmarr}) of the Smarr relation, as has been commonly used in work on Lifshitz black holes.  
Our approach 
for obtaining mass agrees in all cases with other methods for black hole mass
when a sufficient number of length scales are present to remove any ambiguity.

In cases where the length scale from the cosmological constant and the horizon radius are not independent, 
an ambiguity arises that can be dealt with in various ways.  
The most challenging examples are the Lifshitz black holes with a Proca field, in which a mass parameter is absent.  
We can use a fictitious mass to specify thermodynamic values which satisfy our Smarr and the first law of thermodynamics. 
This method yields a family of results, of which we may choose one given an independent derivation.
We also attempted to find the thermodynamic mass and volume by using analogies between the charged Lifshitz black hole solution and RN-AdS black hole.  
For $k=1$ and $k= -1$ our thermodynamic analysis yields zero electric potential at the horizon despite the presence of charge.
For the $k=0$ case, there is no mass parameter; assuming independence of the length scales we obtain a non-vanishing electric potential $\Phi$. It  remains an open question as to how this $\Phi$ can be explained from $A_{t}$, which grows as $r^{2}$.

We also found that the reverse isoperimetric inequality \cite{Cvetic2010} does not hold in general for all of
the Lifshitz cases ($z>1$) we studied, for at least some values of horizon radius. In this sense
Lifshitz black holes are also `super-entropic' -- their entropy is larger than their thermodynamic volume
would na\"ively allow --  a phenomenon recently observed for a new
class of ultraspinning black holes \cite{Hennigar2014}.   
The reason appears to be that the Lifshitz parameter
modifies the scaling of the thermodynamic volume assuming the same identification of pressure
as the AdS scenario, with  $V \sim r_h^{D+z-2}$.
The necessary and sufficient conditions under which the reverse isoperimetric inequality holds remains
an interesting subject for further study.
 
The most compelling future work will be the use of this technique to obtain the mass
for numerical Lifshitz-symmetric black hole solutions, and to use this to come to a better
understanding of thermodynamics in Lifshitz spacetimes.

Another interesting future study is the application of this method to spacetimes where
the mass does not tend to zero as the horizon radius approaches zero. This is
particularly intriguing in the context of soliton solutions, as it may apply
holographically to the Casimir energy of various field theories.

\acknowledgments We thank Sergey Solodukhin
(CNRS, Tours) for helpful discussions. M. P. would like to thank Sang-Heon Yi
for useful discussions on mass for Lifshitz spacetimes.
This work has been supported by the National
Sciences and Engineering Research Council of Canada. W. B. was
funded by the Vanier CGS Award. M. P. was supported by the National Research Foundation of Korea (NRF) funded
by the Korea government with the Grant No. 2013R1A6A3A01065975 and for the travel grant by Center for Women in Science, Engineering, and Technology (WISET).

\bibliography{jabref}{}
\bibliographystyle{h-physrev}

\end{document}